\documentclass[12pt,a4paper,aps,preprint,superscriptaddress,nofootinbib]{revtex4-1}
\usepackage[utf8]{inputenc}
\usepackage{graphicx}
\usepackage{amssymb}
\usepackage{textcomp}
\usepackage{amsmath}
\usepackage{tabularx}
\usepackage{bm}
\usepackage{times}
\usepackage{color}
\usepackage{slashed}
\usepackage{multirow}
\usepackage{verbatim}
\usepackage{cancel}
\usepackage{subfigure}
\usepackage{ulem}

\usepackage[colorlinks=true, pdfstartview=FitV, linkcolor=blue, citecolor=blue, urlcolor=blue]{hyperref}
\allowdisplaybreaks[4]

\newcommand{\minigraph}[5][0.25in]{\begin{minipage}{#2}\begin{center}\includegraphics[width=#2]{#5}\\\vspace{#3}\hspace{#1}{\footnotesize #4}\end{center}\end{minipage}}

\def\lsim{\mathrel{\raise.3ex\hbox{$<$\kern-.75em\lower1ex\hbox{$\sim$}}}}
\def\gsim{\mathrel{\raise.3ex\hbox{$>$\kern-.75em\lower1ex\hbox{$\sim$}}}}

\def\ev{\,{\rm eV}}

\definecolor{orange}{rgb}{1,0.5,0}

\begin{document}

\title{Probing heavy triplet leptons of the type-III seesaw mechanism at future muon colliders}

\author{Tong Li}
\email{litong@nankai.edu.cn}
\affiliation{
School of Physics, Nankai University, Tianjin 300071, China
}
\author{Han Qin}
\email{han.qin@pitt.edu}
\affiliation{
PITT PACC, Department of Physics and Astronomy, University of Pittsburgh, Pittsburgh, PA 15217, USA
}
\author{Chang-Yuan Yao}
\email{yaocy@nankai.edu.cn}
\affiliation{
School of Physics, Nankai University, Tianjin 300071, China
}
\affiliation{
Deutsches Elektronen-Synchrotron DESY, Notkestra{\ss}e 85, 22607 Hamburg, Germany
}
\author{Man Yuan}
\email{yuanman@mail.nankai.edu.cn}
\affiliation{
School of Physics, Nankai University, Tianjin 300071, China
}

\begin{abstract}
We study the search potential of heavy triplet leptons in Type III Seesaw mechanism at future high-energy and high-luminosity muon colliders.
The impact of up-to-date neutrino oscillation results is taken into account on neutrino parameters and the consequent decay patterns of heavy leptons in Type III Seesaw. The three heavy leptons in the Casas-Ibarra parametrization with diagonal unity matrix are distinguishable in their decay modes.
We consider the pair production of charged triplet leptons $E^+ E^-$ through both $\mu^+\mu^-$ annihilation and vector boson fusion (VBF) processes. The leading-order framework of electroweak parton distribution functions are utilized to calculate the VBF cross sections.
The charged triplet leptons can be probed for masses above $\sim 1$ TeV. The $E^+E^-\to ZZ\ell^+\ell^-$ channel can be further utilized to fully reconstruct the three triplet leptons and distinguish neutrino mass patterns.
The pair production of heavy neutrinos $N$ and the associated production $E^\pm N$ are only induced by VBF processes and lead to lepton-number-violating (LNV) signature. We also study the search potential of LNV processes at future high-energy muon collider with $\sqrt{s}=30$ TeV.
\end{abstract}

\maketitle

%%%%%%%%%%%%%%%%%%%%%%%%%%%%%%%%
\section{Introduction}
\label{sec:Intro}
%%%%%%%%%%%%%%%%%%%%%%%%%%%%%%%%

Neutrino oscillation experiments have provided compelling evidence that neutrinos have small but finite masses.
It is well-known that, in the context of the Standard Model (SM), the small Majorana neutrino masses can be generated at leading order through a dimension-5 ``Weinberg operator''~\cite{Weinberg:1979sa}
\begin{eqnarray}
{\kappa\over \Lambda}LLHH\;,
\label{weinberg}
\end{eqnarray}
where $L$ and $H$ stand for the SM left-handed lepton doublet and the Higgs doublet, respectively.
The ultraviolet completions of Eq.~(\ref{weinberg}) at tree level~\cite{Ma:1998dn} are only implemented by the Type I~\cite{Minkowski:1977sc,Yanagida:1979as,GellMann:1980vs,Glashow:1979nm,Mohapatra:1979ia,Shrock:1980ct,Schechter:1980gr}, Type II~\cite{Konetschny:1977bn,Cheng:1980qt,Lazarides:1980nt,Schechter:1980gr,Mohapatra:1980yp} and Type III~\cite{Foot:1988aq} Seesaw mechanisms.
As proved in Refs.~\cite{Kersten:2007vk,Moffat:2017feq}, requiring all three light neutrinos to be massless is equivalent to
requiring lepton number to be conserved at all orders in perturbation theory. In other words, in light of non-zero
neutrino masses, lepton
number is nearly conserved in low-scale seesaw models with fermionic singlets.
Thus, for low-scale Type I Seesaw, the observable lepton number violation (LNV) is decoupled in high
energy processes. Any observation of lepton number violation at high-energy colliders implies an extension of
the canonical Type I Seesaw. In particular, the Type III Seesaw introduces at least two SU$(2)_L$ triplet lepton representations $\Sigma_L$ with $(1, 3, 0)$ SM quantum numbers. The active neutrino mass matrix has the same form as that in Type I Seesaw and the high energy scale $\Lambda$ is given by integrating out the mass of triplet leptons which can be as low as a few TeV.

On the other hand, several reactor experiments have provided precise measurements for the neutrino mixing parameters. For instance, Double Chooz~\cite{DoubleChooz:2011ymz}, RENO~\cite{RENO:2012mkc} and in particular Daya Bay~\cite{DayaBay:2012fng} have reported non-zero measurements of $\theta_{13}$ by looking for the disappearance of anti-electron
neutrino. More recently, two long-baseline neutrino experiments T2K~\cite{T2K:2021xwb} and NOvA~\cite{NOvA:2021nfi} reported the indication of a non-zero CP-violating phase. Despite a slight tension at $\sim 2\sigma$ level between them, these experiments provide us up-to-date neutrino oscillation results to investigate the impact on neutrino mass models.
As the sign of the mass-squared difference $\Delta m_{31}^2$ has not been determined up to now, the so-called normal hierarchy (NH) with positive $\Delta m_{31}^2$ and the inverted hierarchy (IH) with negative $\Delta m_{31}^2$ are normally considered in the data analysis.
Future neutrino oscillation experiments, such as T2HK~\cite{Hyper-KamiokandeProto-:2015xww}, JUNO~\cite{JUNO:2015sjr} and DUNE~\cite{DUNE:2015lol}, are very likely to further explain the discrepancy and make a distinction between these two neutrino mass patterns.
Besides the neutrino experiments, high-energy colliders can also provide complementary strategies to reveal the neutrino properties by searching for the triplet leptons and leptonic signatures.

The new triplet leptons in TeV Type III Seesaw mechanism can be experimentally accessible at the Large Hadron Collider (LHC) (see Refs.~\cite{Arhrib:2009mz,Li:2009mw,Biggio:2011ja,Bandyopadhyay:2011aa,Aguilar-Saavedra:2013twa,Cai:2017mow,Jana:2020qzn,Das:2020uer,Ashanujjaman:2021jhi,Sen:2021fha,Ashanujjaman:2021zrh} and the references therein) and lepton colliders~\cite{Goswami:2017jqs,Das:2020gnt,Bandyopadhyay:2020mnp,Bhattacharya:2021ltd}. The triplet leptons are produced in pairs through pure electroweak (EW) gauge couplings and followed by multi-lepton signatures. After combining two-, three- and four-lepton channels together, the observed lower
limit on the mass of heavy leptons is above 900 GeV at the 95\% confidence level (CL)~\cite{CMS:2017ybg,ATLAS:2022yhd}. According to the theoretical estimations, it is difficult to discover the lepton-number-conserving or lepton-number-violating signal events for triplet leptons heavier than $\sim 1$ TeV at the LHC. One needs a bigger machine with higher energy and integrated luminosity. Recently, high energy muon colliders have received much attention in the community due to the technological developments~\cite{MICE:2019jkl,Delahaye:2019omf,Bartosik:2020xwr}. The community considers the high-energy option of muon collider with multi-TeV center-of-mass (c.m.) energies and a high integrated luminosity scaling with energy quadratically~\cite{Delahaye:2019omf}
\begin{eqnarray}
\label{eq:lumi}
  \mathcal{L}=\left(\frac{\sqrt{s}}{10\ {\rm TeV}}\right)^2 10~\textrm{ab}^{-1}\;.
\end{eqnarray}
It thus provides excellent opportunities to produce and discover new heavy EW particles such as the triplet leptons in Type III Seesaw.

In this work, we investigate the search for heavy triplet leptons at high-energy muon colliders. The pair productions of heavy triplet leptons are categorized into the topologies of direct $\mu^+\mu^-$ annihilation and the vector boson fusion (VBF). The mass eigenstates of charged triplet lepton (denoted by $E^\pm$) can be produced in pairs through both of these two processes. The large total cross section of $E^+E^-$ production makes it an appealing discovery channel of the triplet lepton. By contrast, the associated production of $E^\pm$ and the neutral heavy neutrino $N$ as well as the pair production of $N$ are only induced by VBF processes. In spite of their small production cross sections, they can produce distinctively rare LNV signal and play as crucial channels to determine Majorana neutrino property and distinguish triplet mass eigenstates at high energies.

This paper is organized as follows. In Sec.~\ref{sec:TypeIII}, we first outline the Type III Seesaw model and discuss the constraints from neutrino oscillation experiments on the decay patterns of heavy triplet leptons.
Then we simulate the pair production of heavy triplet leptons via $\mu^+\mu^-$ annihilation and VBF at high-energy muon colliders in Sec.~\ref{sec:Pro}.
The results of projected sensitivity for heavy leptons are also given.
Finally, in Sec.~\ref{sec:Con} we summarize our conclusions.

%%%%%%%%%%%%%%%%%%%%%%%%%%%%%%%%%%%%%%%%%%%%%%
\section{Type III Seesaw and heavy triplet leptons}
\label{sec:TypeIII}
%%%%%%%%%%%%%%%%%%%%%%%%%%%%%%%%%%%%%%%%%%%%%%

In addition to the SM fields, the Type III Seesaw~\cite{Foot:1988aq} consists of SU$(2)_L$ triplet leptons with zero hypercharge, i.e., $\Sigma_L\sim (1,3,0)$ under SM gauge group SU$(3)_C\times$SU$(2)_L\times$U$(1)_Y$:
\begin{eqnarray}
&&\Sigma_L = \left(
  \begin{array}{cc}
    \Sigma_L^0/\sqrt{2} & \Sigma^+_L \\
    \Sigma_L^- & -\Sigma_L^0/\sqrt{2} \\
  \end{array}
\right) \ .
\end{eqnarray}
The Type III Seesaw Lagrangian is given by the kinetic term, mass term and Yukawa interactions in terms of $\Sigma_L$ and its charge conjugated form $\Sigma_L^c$
\begin{equation}
 \mathcal{L}_{T} =
 {\rm Tr}\left[\overline{\Sigma_L}i\cancel{D}\Sigma_L\right]
-\left({1\over 2}{\rm Tr}\left[\overline{\Sigma_L^c}M_\Sigma \Sigma_L\right]+Y_\Sigma \overline{l_L}\tilde{H}\Sigma_L^c+\text{H.c.}\right)\;,
\label{eq:typeIIIMassLag}
\end{equation}
where $\tilde{H}=i\sigma_2 H^\ast\sim (1,2,-1)$.
Besides the neutrino mass mixing for $(\nu_L,\Sigma_L^0)^T$, the Yukawa term in Eq.~(\ref{eq:typeIIIMassLag}) also induces a mass mixing between the charged SM leptons and the charged triplet leptons
\begin{eqnarray}
\left(
  \begin{array}{cc}
    \overline{\nu_L^c} & \overline{\Sigma_L^{0c}} \\
  \end{array}
\right) \left(
  \begin{array}{cc}
    0 & Y_\Sigma^Tv_0/2\sqrt{2} \\
    Y_\Sigma v_0/2\sqrt{2} & M_\Sigma/2 \\
  \end{array}
\right) \left(
  \begin{array}{c}
    \nu_L \\
    \Sigma_L^0 \\
  \end{array}
\right)
+
\left(
  \begin{array}{cc}
    \overline{l_R} & \overline{\Sigma_L^{+c}} \\
  \end{array}
\right) \left(
  \begin{array}{cc}
    m_l & 0 \\
    Y_\Sigma v_0 & M_\Sigma \\
  \end{array}
\right) \left(
  \begin{array}{c}
    l_L \\
    \Sigma_L^- \\
  \end{array}
\right)+\text{H.c.}\;,\nonumber \\
\end{eqnarray}
where $v_0\approx 246$ GeV is the vacuum expectation value of Higgs.
After introducing unitary matrices to transit light doublet and heavy triplet lepton fields, one obtains the diagonal mass matrices and mass eigenvalues for neutrinos and charged leptons. The mass eigenstates of heavy neutral and charged leptons are denoted by $N$ and $E^\pm$, respectively.
Expanding the unitary matrices to leading order in $Y_\Sigma v_0 M_\Sigma^{-1}$, the mixing between the SM charged leptons and triplet lepton mass eigenstates is governed by $V_{\ell N}=-Y_\Sigma^\dagger v_0 M_\Sigma^{-1}/\sqrt{2}$.
In the mass basis, to the leading order, we have the following interactions involving heavy triplet leptons
\begin{eqnarray}
\mathcal{L}_{\rm Type~III}&=&e\overline{E}\gamma^\mu E A_\mu + g \cos\theta_W\overline{E}\gamma^\mu E Z_\mu \nonumber\\
&+&{g\over
2\cos\theta_W}\Big[\overline{\nu_\ell}(V_{PMNS}^\dagger V_{\ell N}\gamma^\mu
P_L-V_{PMNS}^TV^{\ast}_{\ell N}\gamma^\mu P_R)N_{}
+\sqrt{2}\overline{\ell}V_{\ell N}\gamma^\mu
P_LE_{}\Big]Z_\mu \nonumber\\
&-&g \Big[ \overline{E}\gamma^\mu N
+{1\over \sqrt{2}}\overline{\ell}V_{\ell N}\gamma^\mu
P_LN_{} + \overline{E}V_{\ell N}^TV_{PMNS}^\ast\gamma^\mu
P_R\nu_{\ell}\Big]W^-_\mu \nonumber\\
&+&{g\over 2M_W}\Big[\overline{\nu_\ell}(V_{PMNS}^\dagger
V_{\ell N}M_{N}^{diag}P_R+V_{PMNS}^TV^{\ast}_{\ell N}M_{N}^{diag}P_L)N_{}+
\sqrt{2}\overline{\ell}V_{\ell N}M_{E}^{diag}P_RE_{}\Big]h + \text{H.c.}\;,\nonumber\\
\label{eq:L}
\end{eqnarray}
where all fields are mass eigenstates.

The above gauge interactions between SM leptons and heavy triplet leptons are all given by $V_{\ell N}$~\cite{Aguilar-Saavedra:2009fxa,Li:2009mw} and the partial widths of both heavy charged lepton and heavy neutrino are proportional to $|V_{\ell N}|^2$.
In the limit of $M_E\approx M_N\gg M_W, M_Z, M_{h}$, the partial widths become~\cite{delAguila:2008cj,Aguilar-Saavedra:2009fxa}
\begin{eqnarray}
&&{1\over 2}\Gamma(N\to \sum_\ell \ell^+ W^-+\ell^-W^+)\approx \Gamma(N\to \sum_\nu\nu Z+\bar{\nu}Z)\approx \Gamma(N\to \sum_\nu \nu h+\bar{\nu}h)\nonumber \\
&\approx &{1\over 2} \Gamma(E^\pm\to \sum_\nu \overset{(-)}\nu W^\pm)\approx \Gamma(E^\pm\to \sum_\ell \ell^\pm Z)\approx \Gamma(E^\pm\to \sum_\ell \ell^\pm h)
\nonumber\\
&\approx & {G_F\over 8\sqrt{2}\pi}\sum_\ell |V_{\ell N}|^2 M_\Sigma^3\;.
\end{eqnarray}
Thus, the decay branching ratios of heavy leptons exhibit
asymptotic behavior consistent with the Goldstone Equivalence Theorem~\cite{Chanowitz:1985hj,Lee:1977yc},
and are given by the following relations~\cite{Franceschini:2008pz,delAguila:2008cj,Aguilar-Saavedra:2009fxa,Arhrib:2009mz}
\begin{eqnarray}
&&{1\over 2}{\rm BR}(N\to \sum_\ell \ell^+ W^-+\ell^-W^+)\approx {\rm BR}(N\to \sum_\nu\nu Z+\bar{\nu}Z)\approx {\rm BR}(N\to \sum_\nu \nu h+\bar{\nu}h)\nonumber \\
&\approx &{1\over 2} {\rm BR}(E^\pm\to \sum_\nu \overset{(-)}\nu W^\pm)\approx {\rm BR}(E^\pm\to \sum_\ell \ell^\pm Z)\approx {\rm BR}(E^\pm\to \sum_\ell \ell^\pm h)
\approx {1\over 4}\;.
\label{eq:BRrelation}
\end{eqnarray}
As the triplet mass grows, this asymptotic behavior can be seen explicitly in the triplet lepton partial widths.

There is a relation between diagonalized neutrino mass matrices and mixing matrices
\begin{eqnarray}
U^\ast_{\rm PMNS}m_\nu^{diag} U^\dagger_{\rm PMNS}+V^\ast_{\ell
N}M_N^{diag}V^\dagger_{\ell N}=0 \ ,
\label{typei}
\end{eqnarray}
with mass eigenvalues being $m_\nu^{diag}=(m_1,m_2,m_3)$ and $M_N^{diag}=(M_{N_1},M_{N_2},M_{N_3})$.
Here the masses and mixing of the light neutrinos in the first term are measurable from the neutrino oscillation experiments. The second term contains the masses and mixing of the new heavy neutrinos.
The general solution of the $V_{\ell N}$ in Eq.~(\ref{typei}) can be parameterized in terms of an arbitrary orthogonal complex matrix $\Omega$ in the Casas-Ibarra parametrization~\cite{Casas:2001sr}
\begin{eqnarray}
V_{\ell N}=U_{\rm PMNS} (m_\nu^{diag})^{1/2}\Omega (M_N^{diag})^{-1/2},
\label{omega}
\end{eqnarray}
with the orthogonality condition $\Omega \Omega^T=I$.
Using the SM electroweak current for heavy Majorana neutrinos $N_i$, in the mixed mass-flavor basis, one can obtain the partial width of their decay into charged lepton~\cite{Atre:2009rg}
\begin{eqnarray}
\Gamma(N_i\to \ell^\pm W^\mp)={G_F\over 8\sqrt{2}\pi}|V_{\ell N_i}|^2 M_{N_i}(M_{N_i}^2+2M_W^2)\left(1-{M_W^2\over M_{N_i}^2}\right)^2,
\label{partialN}
\end{eqnarray}
where $\ell=e,\mu,\tau$.
Note that generally the Dirac mass matrix can be quite arbitrary with three complex angles parameterizing the orthogonal matrix $\Omega$~\cite{FileviezPerez:2009hdc,Biggio:2019eeo,Das:2020uer,FileviezPerez:2020cgn}. The above asymptotic behavior of decay branching ratios of heavy neutrino indicates BR$(N_i\to \sum_\ell \ell^\pm W^\mp)$ is nearly equal to $25\%$ for any heavy neutrino $N_i~(i=1,2,3)$. The branching fraction of each heavy neutrino $N_i$ decaying into any charged lepton flavor is thus less than $25\%$ for any possible $\Omega$ matrix. In one simple case for $\Omega$, i.e., a diagonal unity matrix $\Omega=I$, the values of $|V_{\ell N}|^2$ for each $N_i$ are proportional to one and only one light neutrino mass~\cite{FileviezPerez:2009hdc}. The branching ratio of $N_{i}\to \ell^\pm W^\mp$ for any charged lepton flavor is thus independent of active neutrino mass. This representative choice of $\Omega=I$ can also lead to different leading decay into charged lepton flavor for each heavy neutrino, so one can distinguish different heavy neutrinos~\cite{FileviezPerez:2020cgn,Han:2021pun}.
Thus, to reduce the parameter dependence and obtain certain predictions, we take this illustrative case of Casas-Ibarra parametrization with $\Omega=I$ in the following study. Note that our results below depend on this specific case and a different choice of $\Omega$ would lead to different conclusions for individual $N_i$.

In order to consider the implication of the neutrino experiments,
we then discuss the neutrino mass and mixing parameters in light of oscillation data.
The neutrino mixing matrix can be parameterized as
\begin{eqnarray}
U_{\text{PMNS}}= \left(
\begin{array}{lll}
 c_{12} c_{13} & c_{13} s_{12} & e^{-\text{i$\delta $}} s_{13}
   \\
 -c_{12} s_{13} s_{23} e^{\text{i$\delta $}}-c_{23} s_{12} &
   c_{12} c_{23}-e^{\text{i$\delta $}} s_{12} s_{13} s_{23} &
   c_{13} s_{23} \\
 s_{12} s_{23}-e^{\text{i$\delta $}} c_{12} c_{23} s_{13} &
   -c_{23} s_{12} s_{13} e^{\text{i$\delta $}}-c_{12} s_{23} &
   c_{13} c_{23}
\label{PMNS}
\end{array}
\right)\times \text{diag} (e^{i \Phi_1/2}, 1, e^{i \Phi_2/2})\;,\nonumber \\
\end{eqnarray}
where $s_{ij}\equiv\sin{\theta_{ij}}$, $c_{ij}\equiv\cos{\theta_{ij}}$, $0 \le
\theta_{ij} \le \pi/2$ and $0 \le \delta, \Phi_i \le 2\pi$ with $\delta$ being the Dirac CP phase and $\Phi_i$ are the Majorana phases.
Taking into account the atmospheric neutrino data from the Super-Kamiokande collaboration,
the latest best global fit results of the neutrino masses and mixing parameters are given by~\cite{Esteban:2020cvm,nufit2021}
\begin{eqnarray}
\Delta m_{21}^2 &=&  7.42 \times 10^{-5} \ev^2 \;, \nonumber \\
\Delta m_{31}^2 &=&  2.510 \times 10^{-3} \ev^2 \ (\Delta m_{32}^2  =  -2.490 \times 10^{-3} \ev^2 )\;, \nonumber \\
\sin^2{\theta_{12}} &=&  0.304\;,~~\sin^2{\theta_{23}} =  0.450 \ (0.570)\;, \nonumber \\
\sin^2{\theta_{13}} &=& 0.02246 \ (0.02241) \;,~~\delta_{\rm CP} = 230^\circ (278^\circ)\;,
\end{eqnarray}
for the spectrum of the neutrino masses of NH (IH).
In addition, the sum of neutrino masses is constrained to be $\sum_{i=1}^3 m_i < 0.12$ eV
by combining the Planck, lensing and baryon acoustic oscillation data~\cite{Planck:2018vyg} at $95\%$ CL.
By inputting the above experimental constraints to the mixing matrix $V_{\ell N}$ and the above partial decay width formulas, we can obtain the preferred values of decay branching ratios for heavy neutrino decay patterns.
In our numerical calculations below, we adopt the benchmark decay branching ratios of heavy neutrinos with $\Omega=I$ for NH and IH as shown in Table~\ref{BRN}.
One can also obtain the decay branching fractions of charged leptons $E_i^\pm$ as the relations in Eq.~(\ref{eq:BRrelation}) apply.

\begin{table}[tb]
\begin{center}
\begin{tabular}{|c|c|c|c|}
\hline
BR$(N_i)$  & $e^\pm W^\mp$ & $\mu^\pm W^\mp$ & $\tau^\pm W^\mp$
\\ \hline
$N_1$ & 17\% (17\%) & 3.26\% (3.73\%) & 4.74\% (4.27\%)
\\ \hline
$N_2$ & 7.43\% (7.43\%) & 10.75\% (7.34\%)  & 6.82\% (10.23\%)
\\ \hline
$N_3$ & 0.56\% (0.56\%) & 11\% (13.93\%)  & 13.44\% (10.51\%)
\\ \hline
\end{tabular}
\end{center}
\caption{Benchmark decay branching ratios of $N_i$ in $\Omega=I$ case for NH (IH). Here we assume the smallest light neutrino mass as $10^{-4}$ eV and $M_{N_i}\gg M_W$.}
\label{BRN}
\end{table}

%%%%%%%%%%%%%%%%%%%%%%%%%%%%%%%%%%%%%%%%%
\section{Pair production of heavy leptons at muon collider}
\label{sec:Pro}
%%%%%%%%%%%%%%%%%%%%%%%%%%%%%%%%%%%%%%%%%

\begin{figure}[h!]
\begin{center}
\minigraph{5cm}{-0.05in}{(a)}{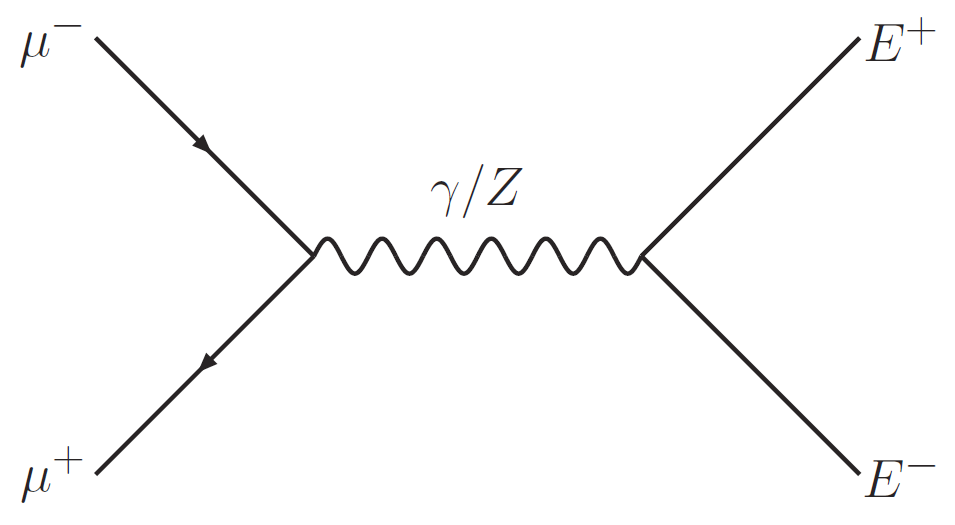}
\minigraph{5cm}{-0.05in}{(b)}{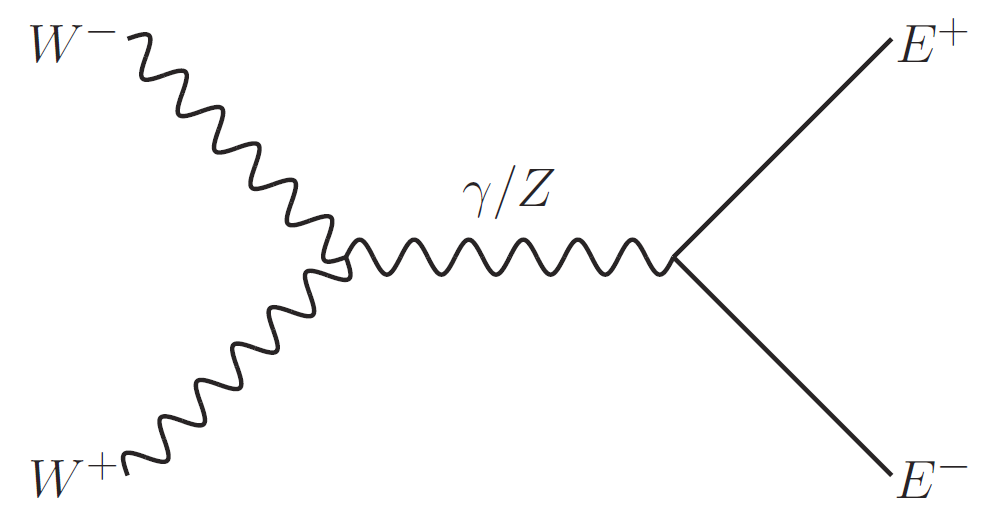}\\
\minigraph{5cm}{-0.05in}{(c)}{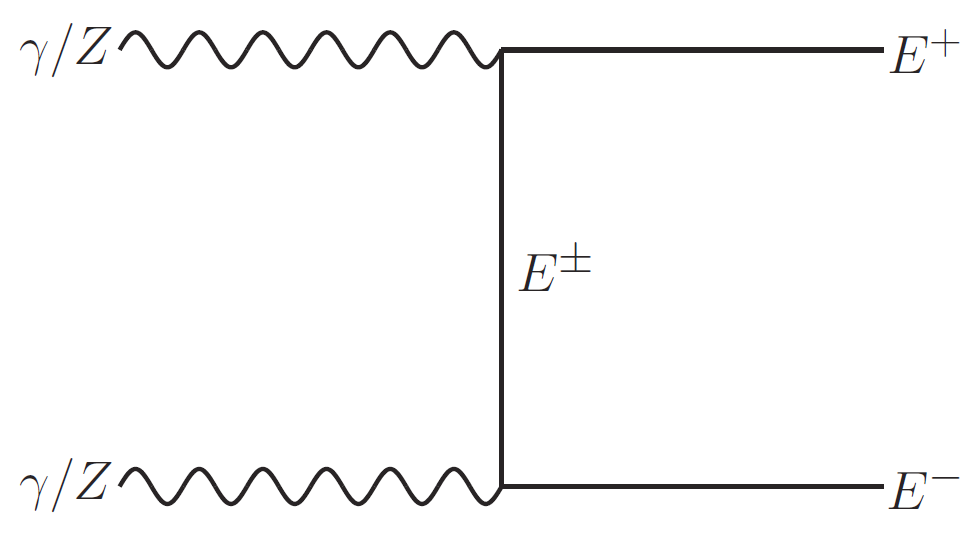}
\minigraph{5cm}{-0.05in}{(d)}{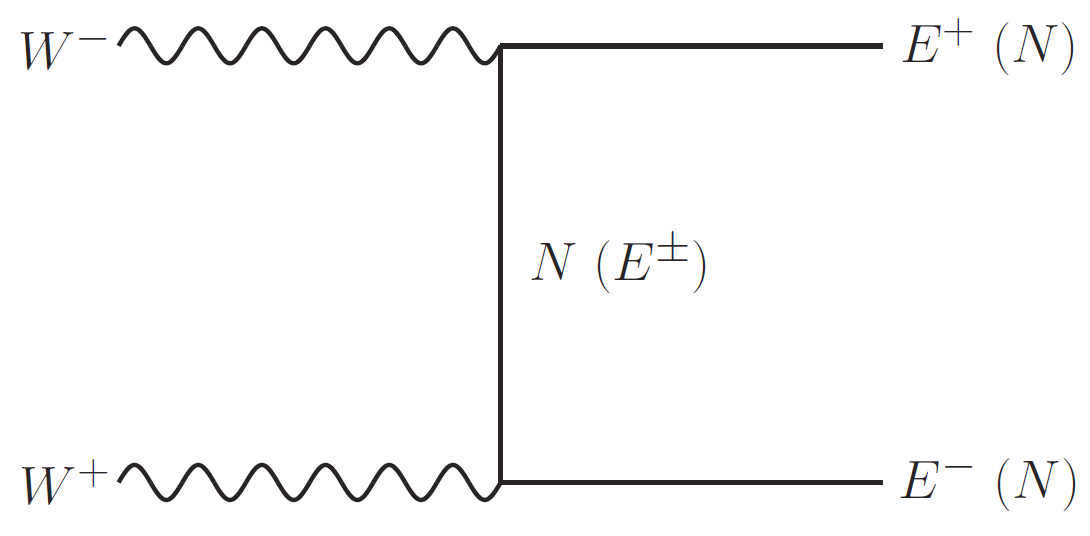}\\
\minigraph{5cm}{-0.05in}{(e)}{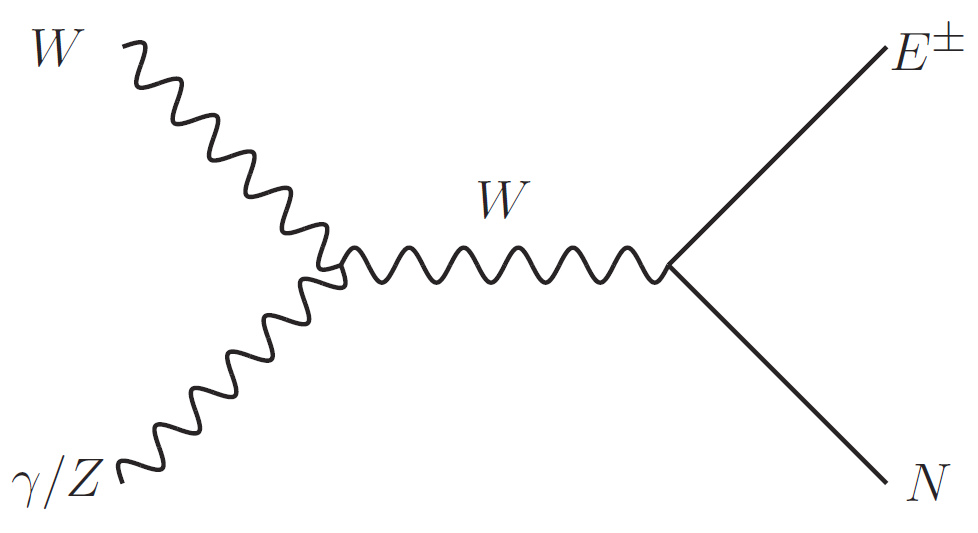}
\minigraph{5cm}{-0.05in}{(f)}{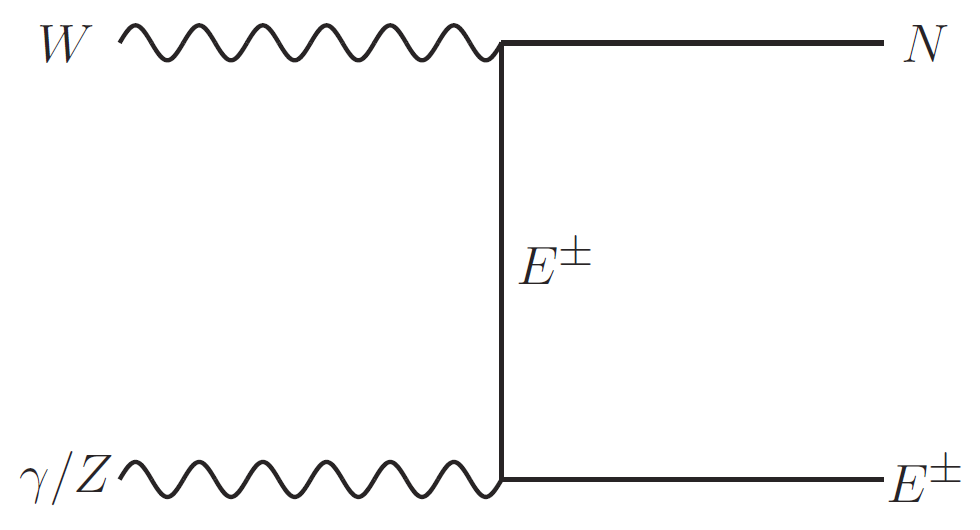}
\end{center}
\caption{Feynman diagrams of heavy triplet lepton pair productions in Type III Seesaw, including $\mu^+\mu^-$ annihilation (a) and VBF processes (b-f).}
\label{diagram:typeIIIpair}
\end{figure}

The pair production channels of heavy triplet leptons at muon colliders include
\begin{itemize}
\item $\mu^+\mu^- \to \gamma^\ast/Z^\ast \to E^+ E^-$\;,
\item $W^+W^- \to \gamma^\ast/Z^\ast \to E^+ E^-$\;,~~$\gamma/Z+\gamma/Z\to E^+E^-$\;,~~$W^+W^-\to NN, E^+E^-$\;,
\item $W^\pm+\gamma/Z\to W^{\pm\ast}\to E^\pm N$,~~$W^\pm+\gamma/Z\to E^\pm N$\;.
\end{itemize}
The Feynman diagrams of heavy triplet lepton pair productions in Type III Seesaw are collected in Fig.~\ref{diagram:typeIIIpair}. They are given by the pure EW gauge interactions in Eq.~(\ref{eq:L}) and not suppressed by the mixing $V_{\ell N}$.
For c.m.~energies $\sqrt{s} \gg M_W$, the initial muon beams substantially radiate EW gauge bosons under an approximately unbroken SM gauge symmetry.
The vector bosons are treated as initial state partons and produce ``inclusive'' processes. We use MadGraph5\_aMC@NLO~\cite{Alwall:2014hca} to calculate the VBF cross sections, by utilizing the leading-order framework of electroweak parton distribution functions (EW PDFs)~\cite{Han:2020uid,AlAli:2021let,Ruiz:2021tdt} with a dynamical scale $Q=\sqrt{\hat{s}}/2$, where $\sqrt{\hat{s}}$ is the partonic c.m.~energy.
For the initial gauge bosons $V_a$ and $V_b$, the VBF production cross section can be factorized as the product of the parton luminosity $d\mathcal{L}_{ab}/d\tau$ and the partonic cross section $\hat{\sigma}$~\cite{Han:2020uid}
\begin{eqnarray}
\sigma(\mu^+\mu^-\to F+X)&=&\int^1_{\tau_0}d\tau \sum_{ab}{d\mathcal{L}_{ab}\over d\tau} \hat{\sigma}(V_a V_b\to F)\;,\nonumber \\
{d\mathcal{L}_{ab}\over d\tau}&=&{1\over 1+\delta_{ab}}\int^1_\tau {d\xi\over \xi}\Big[f_a(\xi,Q^2)f_b({\tau\over \xi},Q^2)+(a\leftrightarrow b)\Big]\;,
\end{eqnarray}
where $F (X)$ denotes an exclusive final state (the underlying remnants), $f_a(\xi,Q^2)$ is the EW PDF for vector $V_a$ with $Q$ being the factorization scale, $\tau_0=m_F^2/s$ and $\tau=\hat{s}/s$. By summing over all gauge boson initial states in Fig.~\ref{diagram:typeIIIpair}, one can obtain the total cross section of the ``inclusive'' VBF production processes.

We consider three benchmark choices of muon collider energies and the corresponding integrated luminosities for illustration
\begin{eqnarray}
\sqrt{s} = 3, \ 10\ {\rm and}\ 30\ {\rm TeV},\quad {\mathcal L} = 1, \ 10\  {\rm and}\ 90\ {\rm ab}^{-1}\;.
\label{eq:para}
\end{eqnarray}
The heavy triplet leptons are followed by the decays into charged leptons or active neutrinos as well as EW gauge bosons. We smeared the parton level events with an energy resolution of $\Delta E/E = 2\%$.
Note that we consider the hadronic decay of EW gauge bosons into two jets. The two jets generally form a fat jet here as the gauge bosons are from heavy triplet decay and thus are highly boosted. We perform the parton showering and detector simulation to identify the fat jets and then reconstruct the vector bosons efficiently. The hadronic decay branching fractions and the efficiencies of the gauge boson reconstruction are included in our numerical analysis. The details of this reconstruction will be described below.
We also take a conservative way to estimate the significance as
\begin{eqnarray}
\mathcal{S}={N_S\over\sqrt{N_S+N_B}}\;,
\end{eqnarray}
where $N_S$ and $N_B$ denote the signal and SM background event numbers, respectively.

%%%%%%%%%%%%%%%%%%%%%%%%%%%%%%%%%%%%%%
\subsection{$E^+ E^-$ Pair Production}
%%%%%%%%%%%%%%%%%%%%%%%%%%%%%%%%%%%%%%

\begin{figure}[htbp]
\makebox[\linewidth][c]{%
\centering
\includegraphics[width=0.5\textwidth]{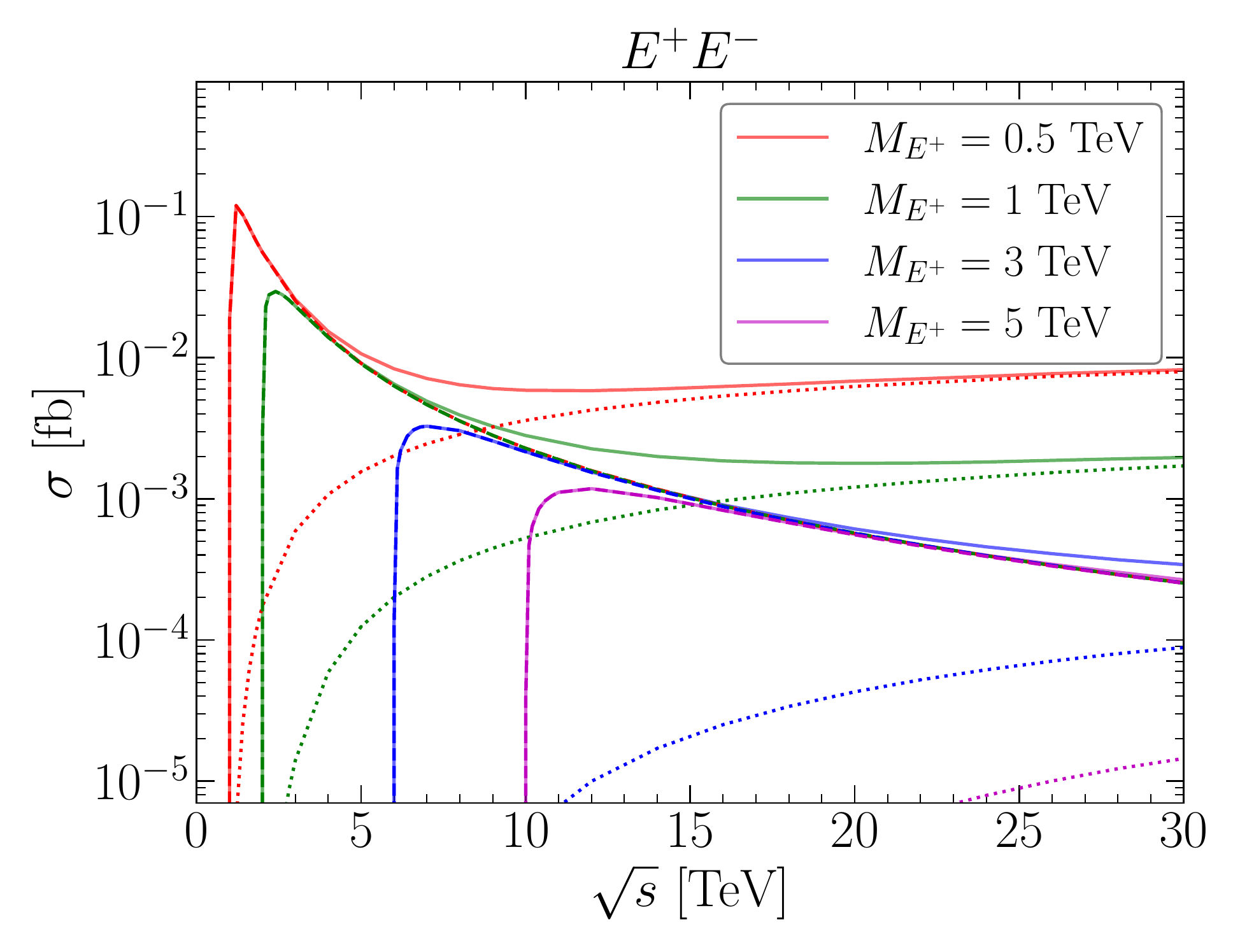}
\includegraphics[width=0.5\textwidth]{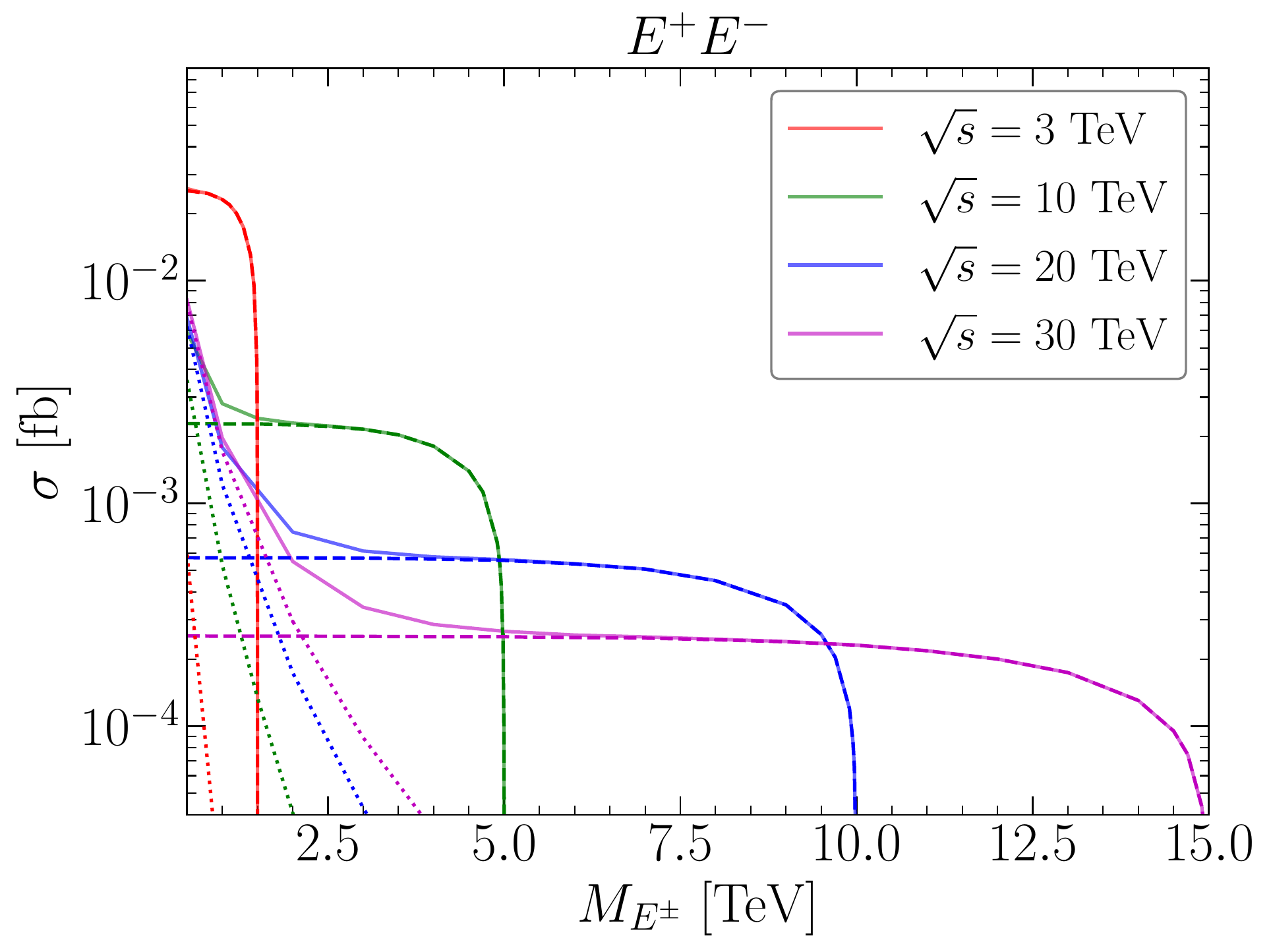}
}
\caption{Cross sections of $E^+ E^-$ pair production as a function of $\sqrt{s}$ (left) and $M_E$ (right) at muon colliders by $\mu^+\mu^-$ annihilation (dashed lines), VBF (dotted lines), and the sum (solid lines).%\CY{the plot label $E^+E^-$ can be removed.}
}
\label{fig:EE_production_sqrts}
\end{figure}

We first perform the calculation of the pair production of heavy charged leptons. The $E^+E^-$ productions are decomposed to $\mu^+ \mu^-$ direct annihilation and the electroweak VBF processes initiated by $\gamma / Z$ or $W^+ W^-$. In Fig.~\ref{fig:EE_production_sqrts}, we show the production cross sections of $E^+E^-$ as a function of $\sqrt{s}$ (left) and $M_E$ (right). The productions via $\mu^+\mu^-$ annihilation, VBF and their sum are represented as dashed lines, dotted lines and solid lines, respectively. The $\mu\mu$ annihilation cross sections approach $\sigma^{\rm Ann}\sim \beta/s$ with the velocity as $\beta=\sqrt{1-\frac{4M_E^2}{s}}$. Thus, they decrease with the increasing c.m. energy. Well above the threshold, the triplet mass can be ignored at high energies and the heavy charged leptons tend to be indistinguishable in this channel. On the other hand, the VBF cross sections $\sigma^{\rm VBF}$ are enhanced by collinear logarithm ${\rm ln}(\hat{s}/m_\mu^2)$ for photon or ${\rm ln}(\hat{s}/m_V^2)$ for massive gauge boson $V$ at high beam energies. They are also very sensitive to the heavy lepton masses and decrease quickly with increasing $M_{E}$. The VBF process is thus important at low $M_{E}$ and high c.m. energies.

We consider the above productions followed by two decay modes of $E^\pm$
\begin{align}
&E^{\pm}\to \nu W^{\pm}\; ,~~E^{\pm}\to \ell^{\pm}Z\;.
\end{align}
For the first mode, the SM neutrinos are summed over and thus one has ${\rm BR}(E^\pm\to \sum_\nu \nu W^\pm)\approx 50\%$ according to the relation in Eq.~(\ref{eq:BRrelation}).
In the second mode, we only study the decay $E^{\pm}\to\mu^{\pm}Z$ for illustration and the decay branching fractions of the heavy neutrino in Table~\ref{BRN} apply here as ${\rm BR}(N_i\to \ell^\pm W^\mp)\approx {\rm BR}(E_i^\pm\to \ell^\pm Z)$ for degenerate triplet leptons. Note that we take the channel $E\to \mu Z$ here for illustrative estimation of the heavy triplet search prospect. The channel $E\to eZ$ can also be analyzed following the same strategies and their SM background analysis is even simpler as the electrons cannot come from the fermion line of initial states.

%%%%%%%%%%%%%%%%%%%%%%%%%%%%%%%%%%%%%%
\subsubsection{$E^+E^-\to W^+W^-\nu\bar{\nu}$}
%%%%%%%%%%%%%%%%%%%%%%%%%%%%%%%%%%%%%%

\begin{figure}[htbp]
\makebox[\linewidth][c]{%
\centering
\includegraphics[width=0.5\textwidth]{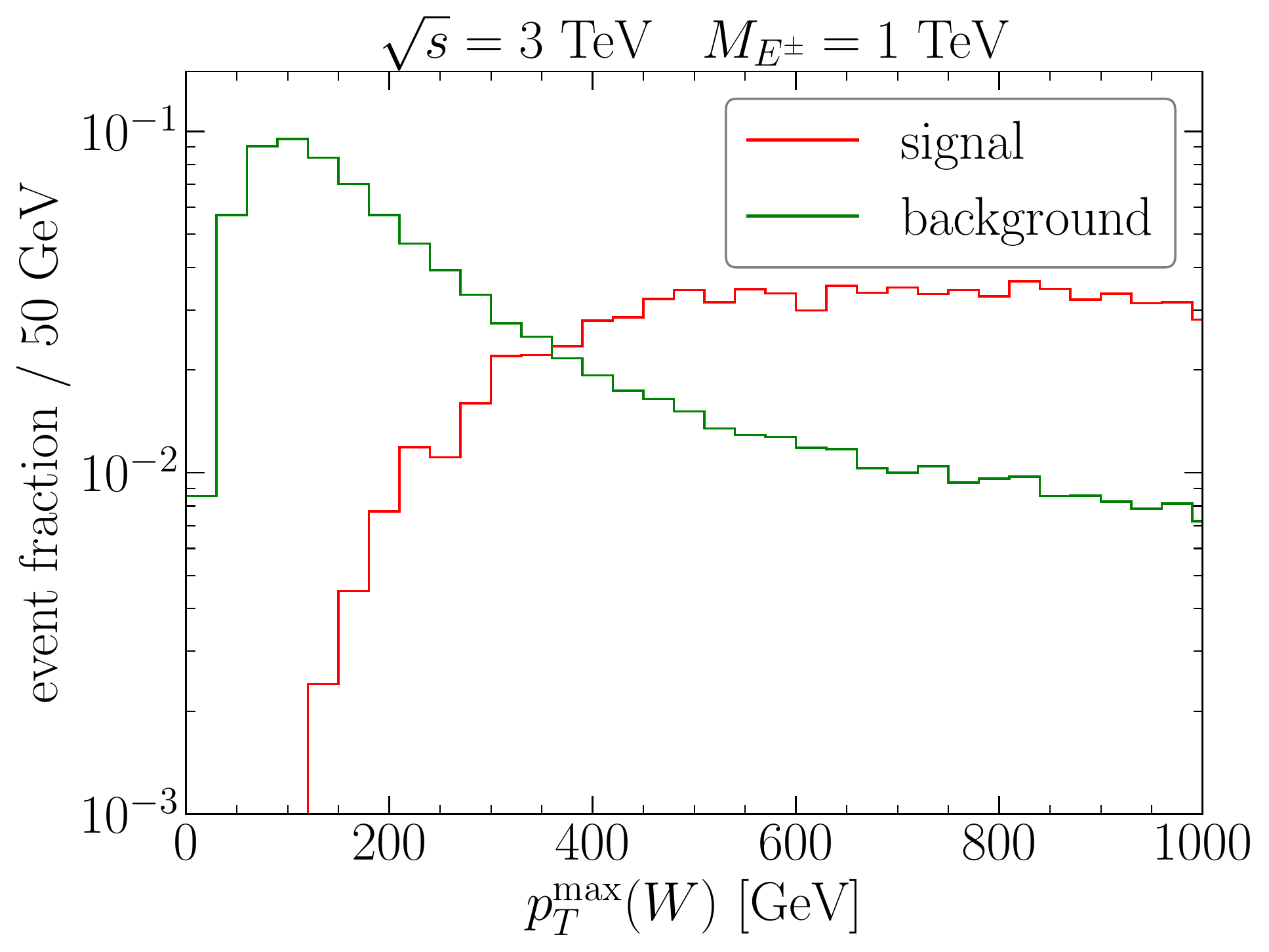}
\includegraphics[width=0.5\textwidth]{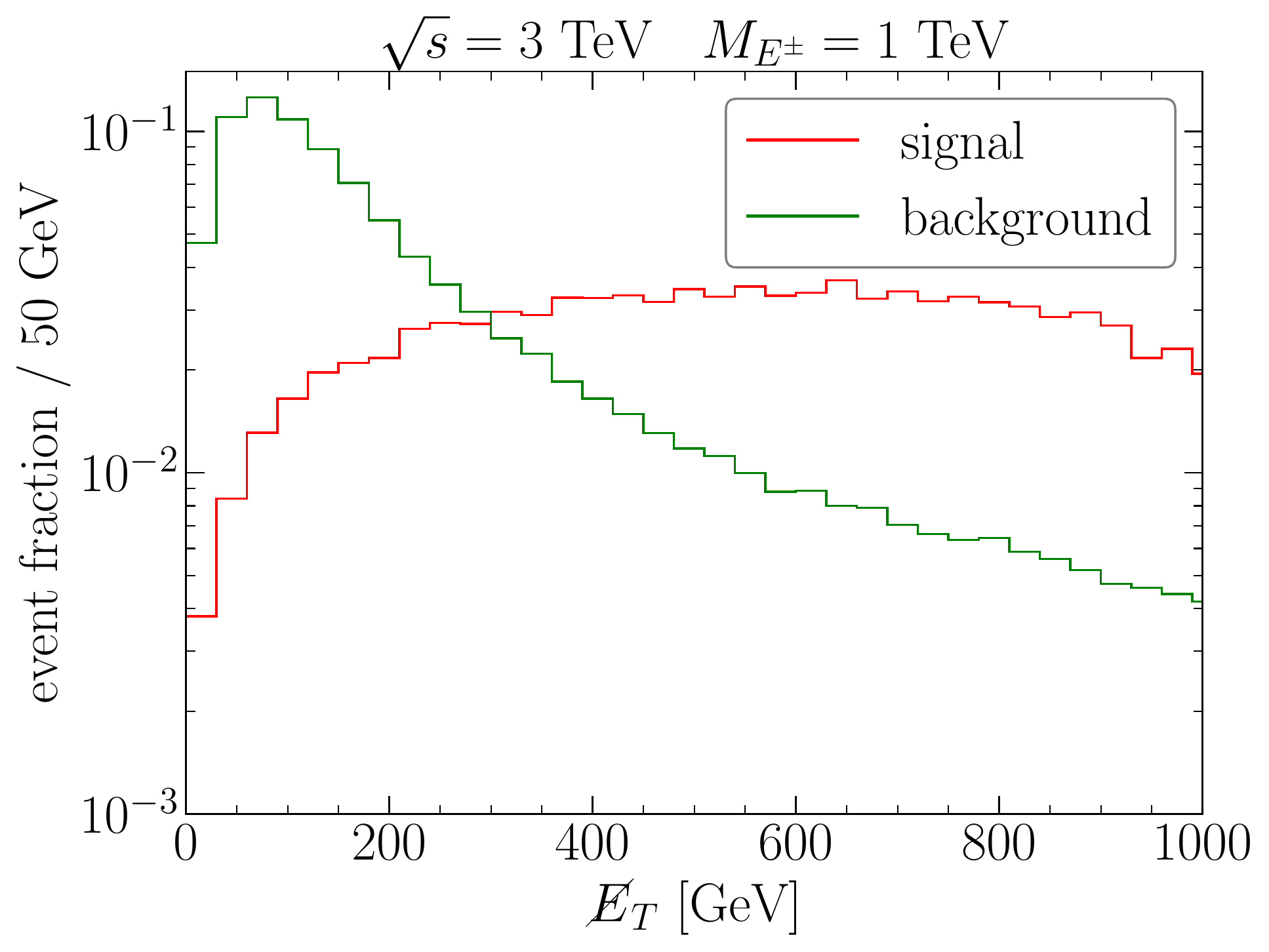}}
\caption{The normalized $p_{T}^{\rm max}(W)$ (left) and $\cancel{E}_{T}$ (right) distributions of the signal $E^+E^-\to W^+W^-\nu \bar{\nu}$ and the SM background for $\sqrt{s}=3$ TeV and $M_E=1$ TeV.
}
\label{fig:Wv-SB-3TeV}
\end{figure}

In this case, our signal is composed of $W^+W^-$ and large missing energy induced by SM neutrinos. The SM background in this channel thus becomes
\begin{eqnarray}
W^+W^-Z\;,
\end{eqnarray}
followed by decay $Z\to \nu\bar{\nu}$ and is generated by MadGraph at the parton level.
We also consider both the $\mu^+\mu^-$ annihilation and VBF productions for the above background. In the beginning, the size of SM background is comparable to our signal.
We thus employ the following judicious cuts to reduce the background:
\begin{itemize}
\item
We take into account the $W$ bosons identified in central region and set the rapidity cut as follows
\begin{eqnarray}
\left| \eta(W) \right|< 3.0 \;.
\end{eqnarray}
\item
For a few TeV $E^{\pm}$, the decay products from heavy triplet lepton are very energetic as shown in Fig.~\ref{fig:Wv-SB-3TeV} (left)~\footnote{The differential cross sections of the $\mu\mu$ annihilation and VBF processes are individually weighted by their total cross sections and summed together here.}. We therefore tight up the kinematical cut with
\begin{eqnarray}
p_{T}^{\rm{max}}(W) > \frac{M_E}{2} \;,
\end{eqnarray}
for the leading $W$ boson.
\item To remove the $ W^+W^-Z$ background with neutrinos from $Z$ boson decay as seen in the right panel of Fig.~\ref{fig:Wv-SB-3TeV}, we require the events with large missing energy from neutrinos
\begin{eqnarray}
\cancel{E}_{T} > \frac{M_E}{4} \;.
\end{eqnarray}
\end{itemize}
As a result, the SM background can be significantly suppressed.

We then perform the local significance analysis. The event numbers of signal ($N_S$) and background ($N_B$) are given by
\begin{eqnarray}
N_S&=&(\sigma_{S}^{\rm Ann} \epsilon_{S}^{\rm Ann}\epsilon_{S,W}^{\rm {Ann}}+\sigma_{S}^{\rm{VBF}} \epsilon_{S}^{\rm{VBF}}\epsilon_{S,W}^{\rm {VBF}})\times {\rm BR}(E^{\pm}\to W^{\pm}\nu)^2 \times \mathcal{L} \times \rm{BR}(W^{\pm} \to q\bar{q}')^2\;, \nonumber \\
N_B&=&(\sigma_{B}^{\rm Ann} \epsilon_{B}^{\rm Ann}\epsilon_{B,W}^{\rm {Ann}}+\sigma_{B}^{\rm{VBF}} \epsilon_{B}^{\rm{VBF}}\epsilon_{B,W}^{\rm {VBF}})\times {\rm BR}(Z\to \nu \bar{\nu})\times \mathcal{L}\times \rm{BR}(W^{\pm} \to q\bar{q}')^2\;,
\label{eqn_N}
\end{eqnarray}
where $\sigma_{S}$ is the signal cross section, $\sigma_{B}$ is the cross section of the SM background $W^+W^-Z$, the superscripts ``${\rm Ann}$'' and ``${\rm VBF}$'' denote $\mu^+\mu^-$ direct annihilation and VBF process, respectively, $\epsilon_{S,B}$ represent the efficiencies of the above cuts and $\mathcal{L}$ denotes the integrated luminosity.
In practise, the $W$ bosons subsequently decay into leptonic or hadronic products. Here and below we consider the hadronic decay of vector bosons into a dijet. As being from heavy lepton decay, the vector bosons are highly boosted and the two jets are likely to merge into a single fat jet $J$. To perform the simulation of vector boson reconstruction, we pass the parton-level events to Pythia 8~\cite{Sjostrand:2014zea} and Delphes 3~\cite{deFavereau:2013fsa} for parton shower and detector simulation, respectively. We choose the detector card of muon collider in the latest version of Delphes for detector-level simulation. The jets from $W$ decay are clustered in FastJet~\cite{Cacciari:2011ma} and we require at least two fat jets. The fat jet is reconstructed via the ``Valencia''
algorithm with $R = 0.5$ and the one with $65 < M_J < 95$ GeV ($75 < M_J < 105$ GeV for $Z$ boson used in Sec.~\ref{sec:EN}) is identified as boosted $W$ candidate. $\epsilon_{S,W}$ and $\epsilon_{B,W}$ in Eq.~(\ref{eqn_N}) are the efficiencies of the vector boson reconstruction.

\begin{figure}[h!]
\begin{center}
\includegraphics[width=0.46\textwidth]{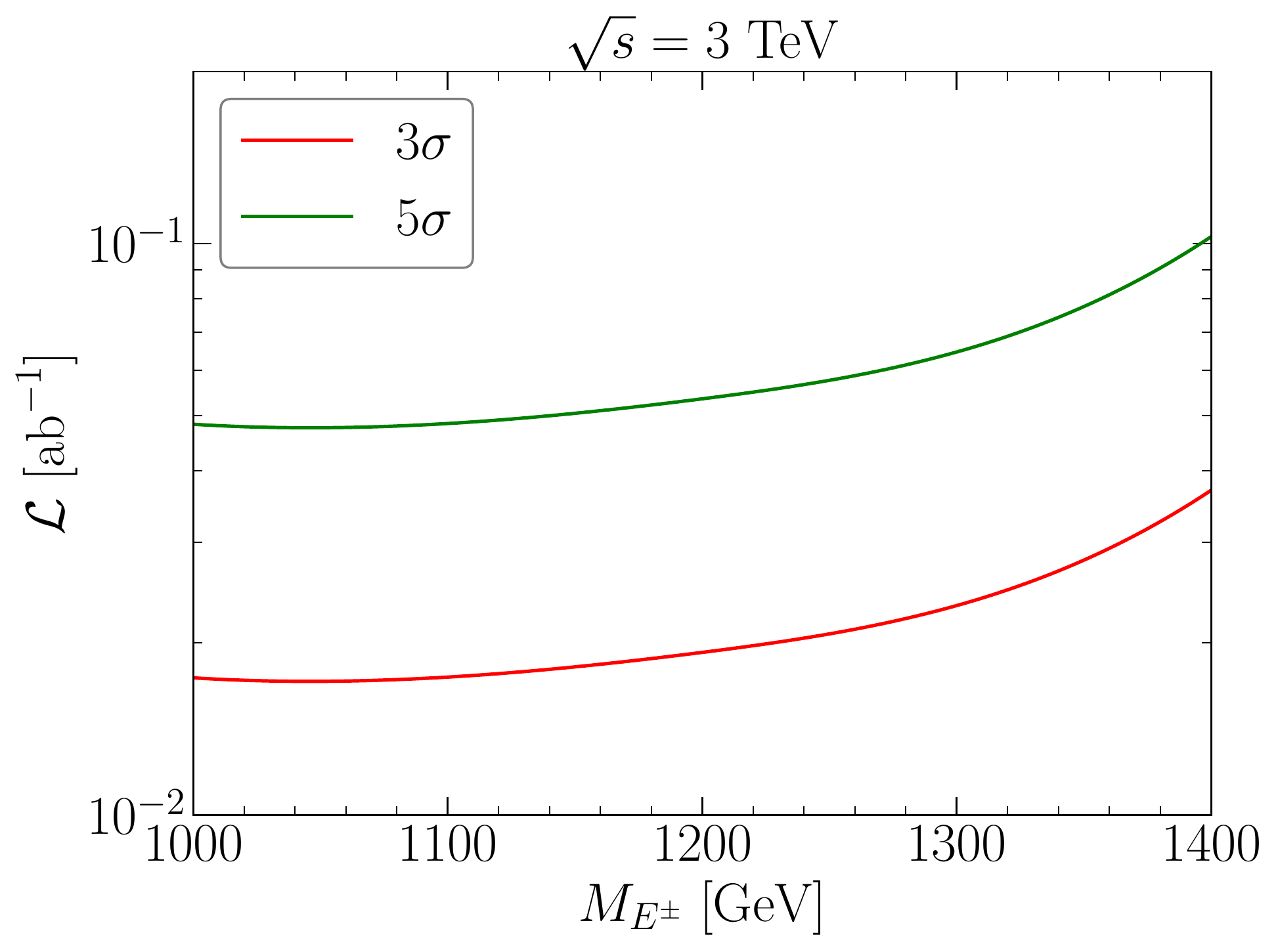}
\includegraphics[width=0.46\textwidth]{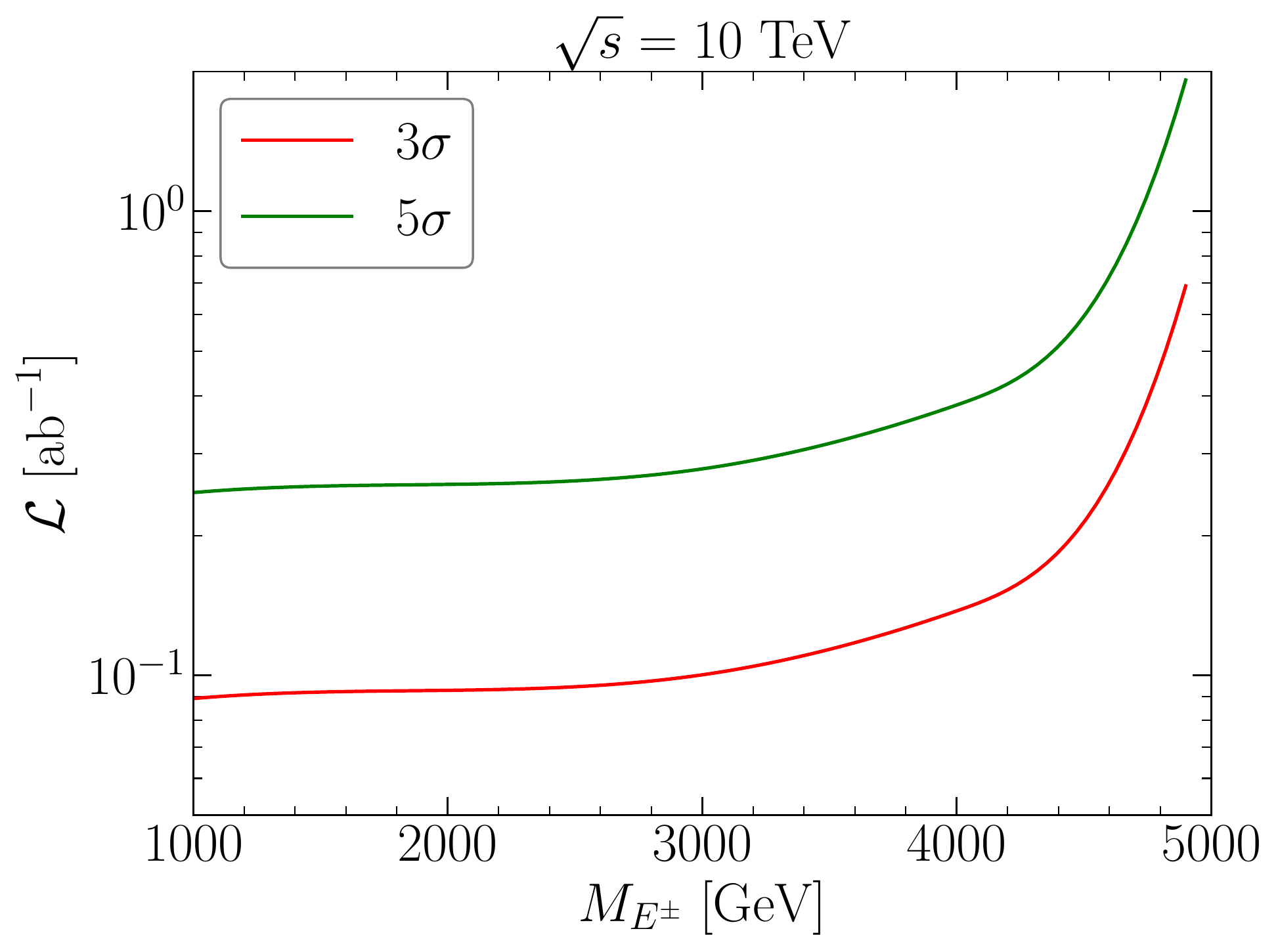}
\end{center}
\caption{The integrated luminosities for $3\sigma$ and $5\sigma$ significance versus $M_E$ for $E^+E^- \to W^+W^-\nu\bar{\nu}$ in muon collider with $\sqrt{s}=3$ TeV (left) and 10 TeV (right).
}
\label{fig:luminosity_EtoWv}
\end{figure}

In Fig.~\ref{fig:luminosity_EtoWv}, we show the integrated luminosities required for $3\sigma$ and $5\sigma$ significance at muon collider with $\sqrt{s}=3$ TeV (left) and $\sqrt{s}=10$ TeV (right). One can see that the charged triplet lepton as heavy as one-half of the colliding energy can be probed. A $5\sigma$ significance can be reached through the $E^+E^-\to W^+W^-\nu\bar{\nu}$ process with the integrated luminosity below 0.1 ab$^{-1}$ for $\sqrt{s}=3$ TeV and 2 ab$^{-1}$ for $\sqrt{s}=10$ TeV.
Despite this optimistic result, however, we are not able to reconstruct the heavy leptons due to the missing neutrinos from $E^\pm$ decay. Moreover, the different neutrino mass patterns and the three triplet leptons cannot be identified and distinguished through this process as the missing neutrinos lose lepton flavor information in detector. We need to discuss other signal channels in the following section to reveal the neutrino characteristics.

%%%%%%%%%%%%%%%%%%%%%%%%%%%%%%%%%%%%%%
\subsubsection{$E^+E^-\to ZZ\mu^+\mu^- $}
%%%%%%%%%%%%%%%%%%%%%%%%%%%%%%%%%%%%%%

\begin{figure}[h!]
\begin{center}
\includegraphics[width=0.46\textwidth]{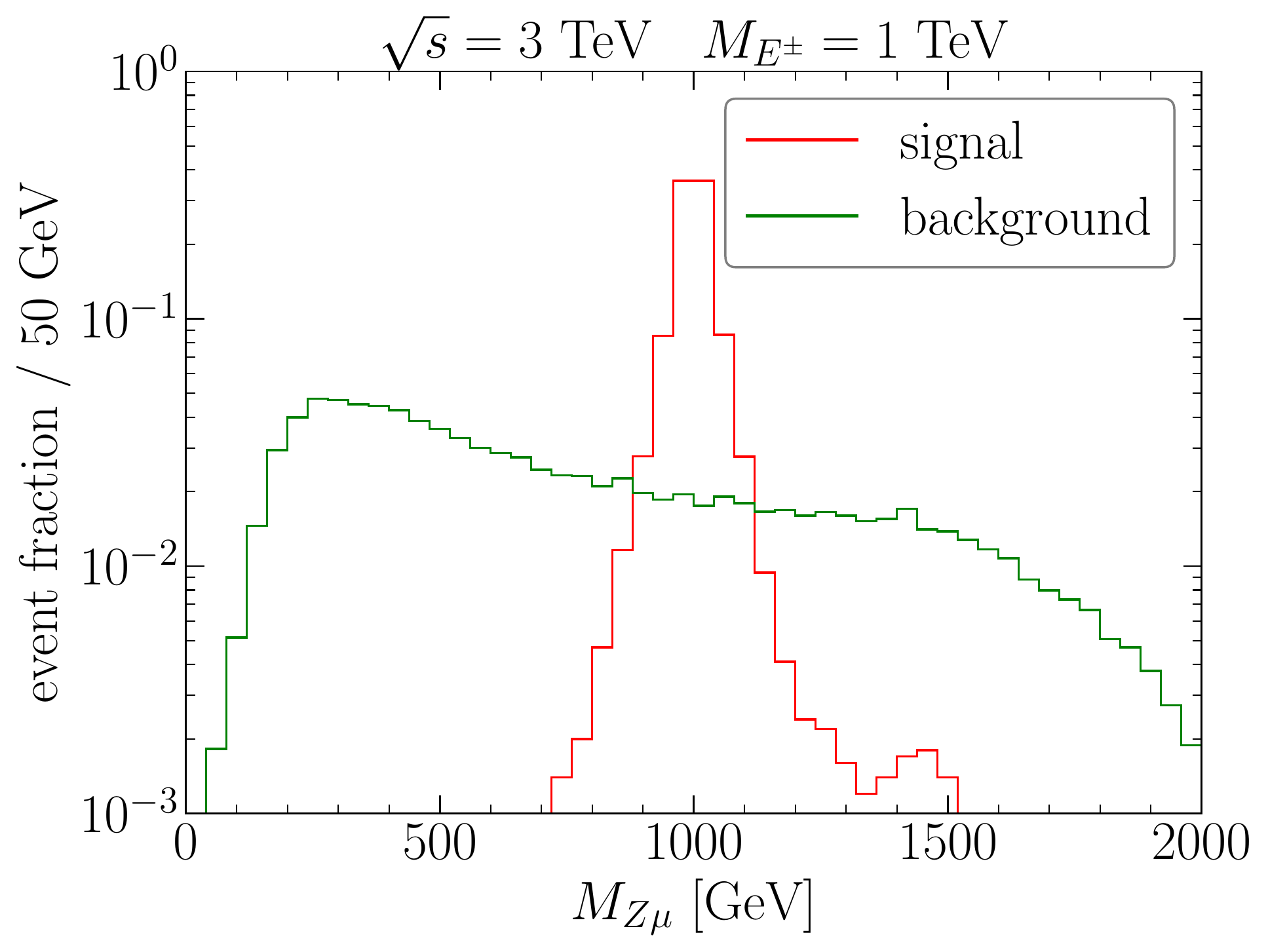}
\includegraphics[width=0.46\textwidth]{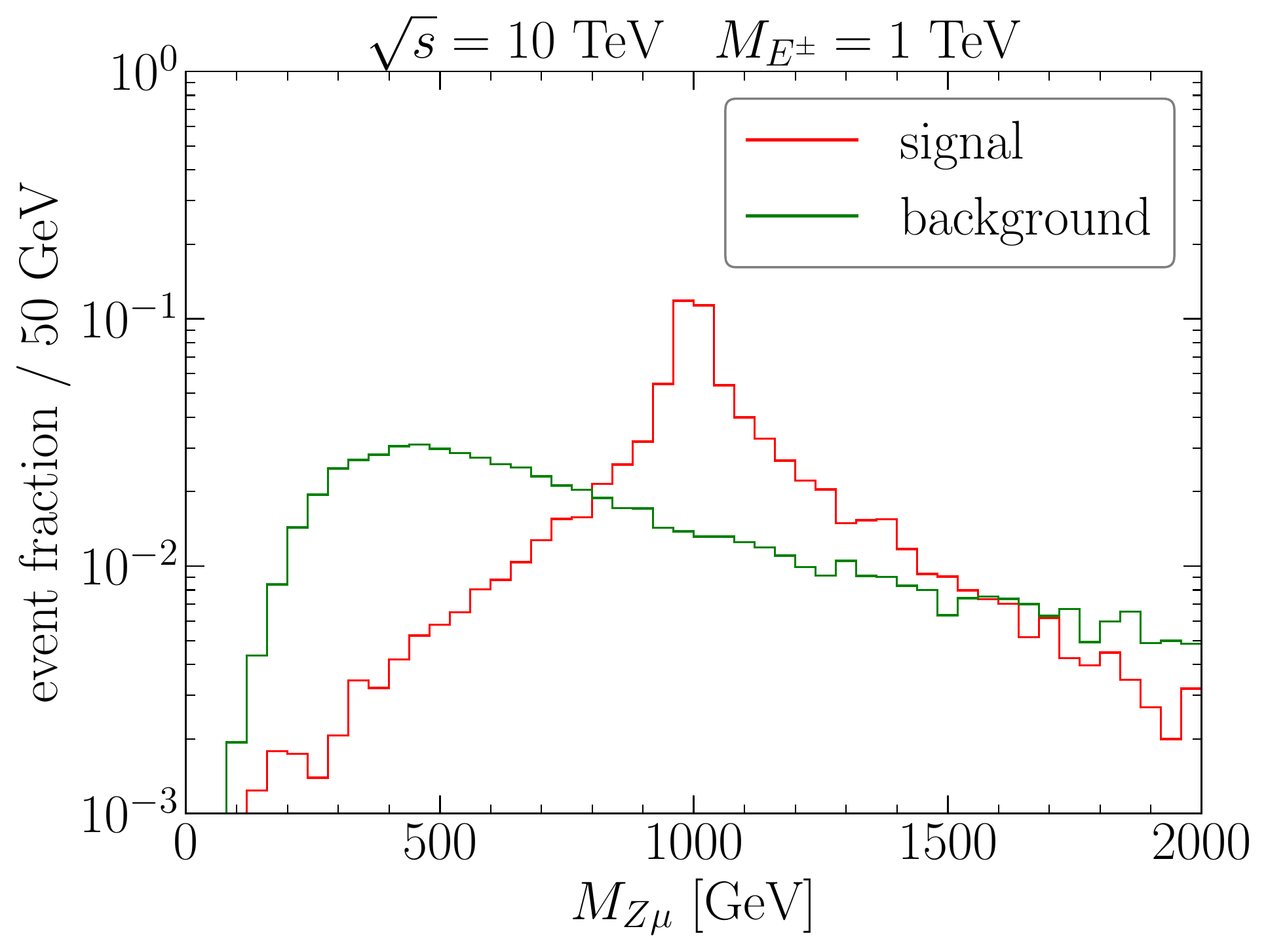}
\end{center}
\caption{The invariant mass $M_{Z\mu}$ distribution of the signal $E^+E^-\to Z Z \mu^+\mu^-$ and the background for $M_E=1$ TeV, $\sqrt{s}=$ 3 TeV (left) and 10 TeV (right).
}
\label{SB-m-muZ}
\end{figure}

Next, we discuss the $E^+E^-$ pair production followed by $E^\pm \to \ell^\pm Z$ decay mode.
In this case, we only consider the decay channel $E^{\pm}\to \mu^{\pm}Z$ for illustration. The signal is thus composed of two $Z$ bosons and $\mu^+\mu^-$.
The SM background is also categorized into $\mu^+\mu^-$ annihilation and VBF production. %They also are generated by MadGraph at the parton level.
Since there are two muons in the final states of this channel, we employ the following basic cuts
\begin{align}
&p_{T}(\mu)>50~{\rm GeV}\, ,~~\left| \eta(\mu, Z) \right|< 3.0\;.
\end{align}
We also employ the judicious cuts to reduce the background:
\begin{itemize}
\item
For the energetic products from a few TeV $E^{\pm}$, we set the following kinematical cuts
\begin{align}
&p_{T}^{\rm max}(Z)>\frac{M_E}{3}\, ,~~p_{T}^{\rm max}(\mu)>\frac{M_E}{3}\;.
\end{align}
\item
Since the final particles $\mu^{\pm}$ and $Z$ are both observable in detector, one can reconstruct the triplet lepton $E^{\pm}$.
To reconstruct heavy fermion $E^{\pm}$, we pair the final particles ($Z,\mu$) by taking advantage of the feature that they have equal masses $M_{Z_1\mu_1}=M_{Z_2\mu_2}$, which is in order to make sure they come from the correct mother particle. Then we take the invariant masses close to $M_E$ with
\begin{align}
\left | M_{Z\mu}-M_E \right| < \frac{M_E}{5}\;.
\end{align}
The reconstructed masses are shown in Fig.~\ref{SB-m-muZ}.
\end{itemize}
These cuts help to reduce the SM backgrounds.

\begin{figure}[h!]
\begin{center}
\includegraphics[width=0.46\textwidth]{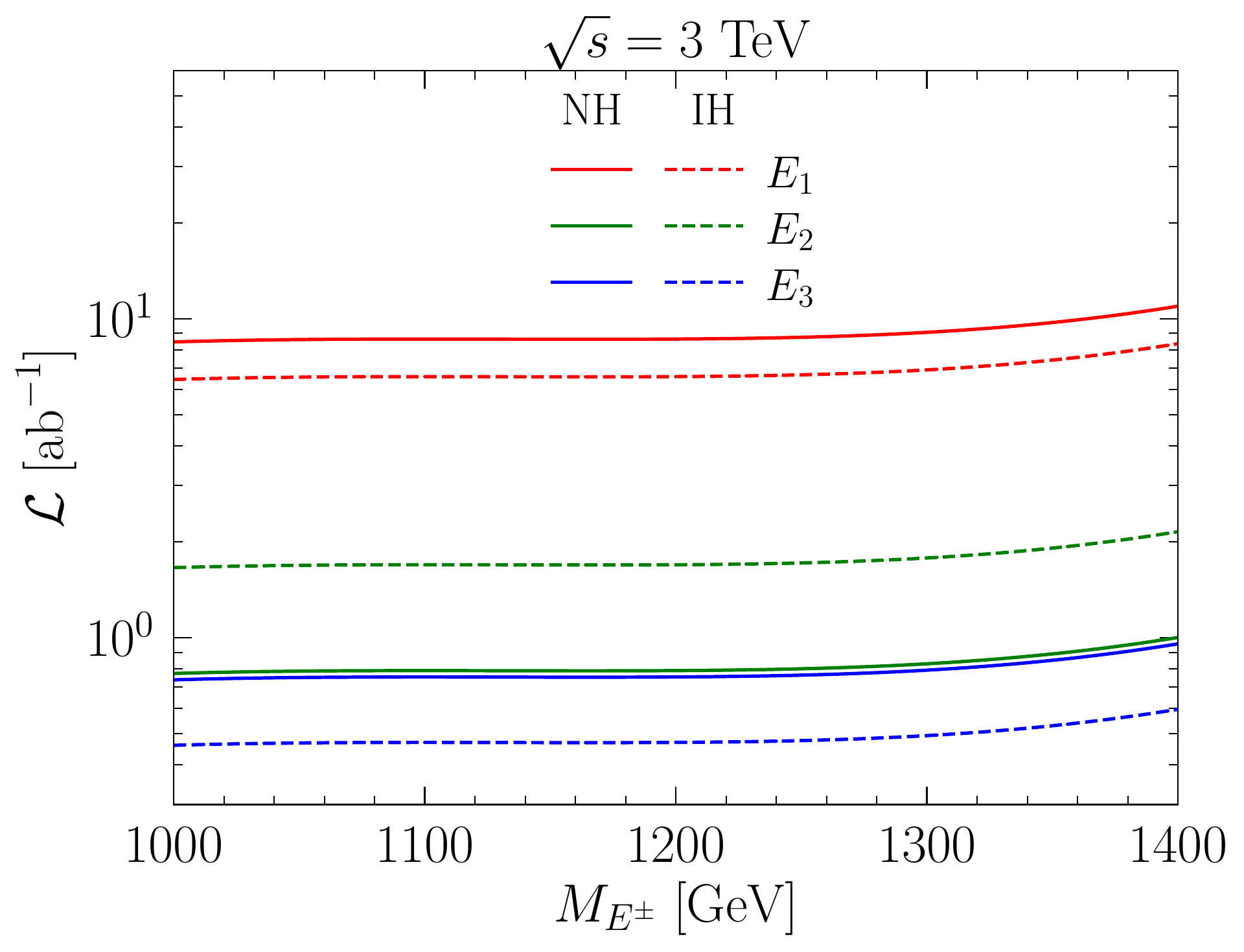}
\includegraphics[width=0.46\textwidth]{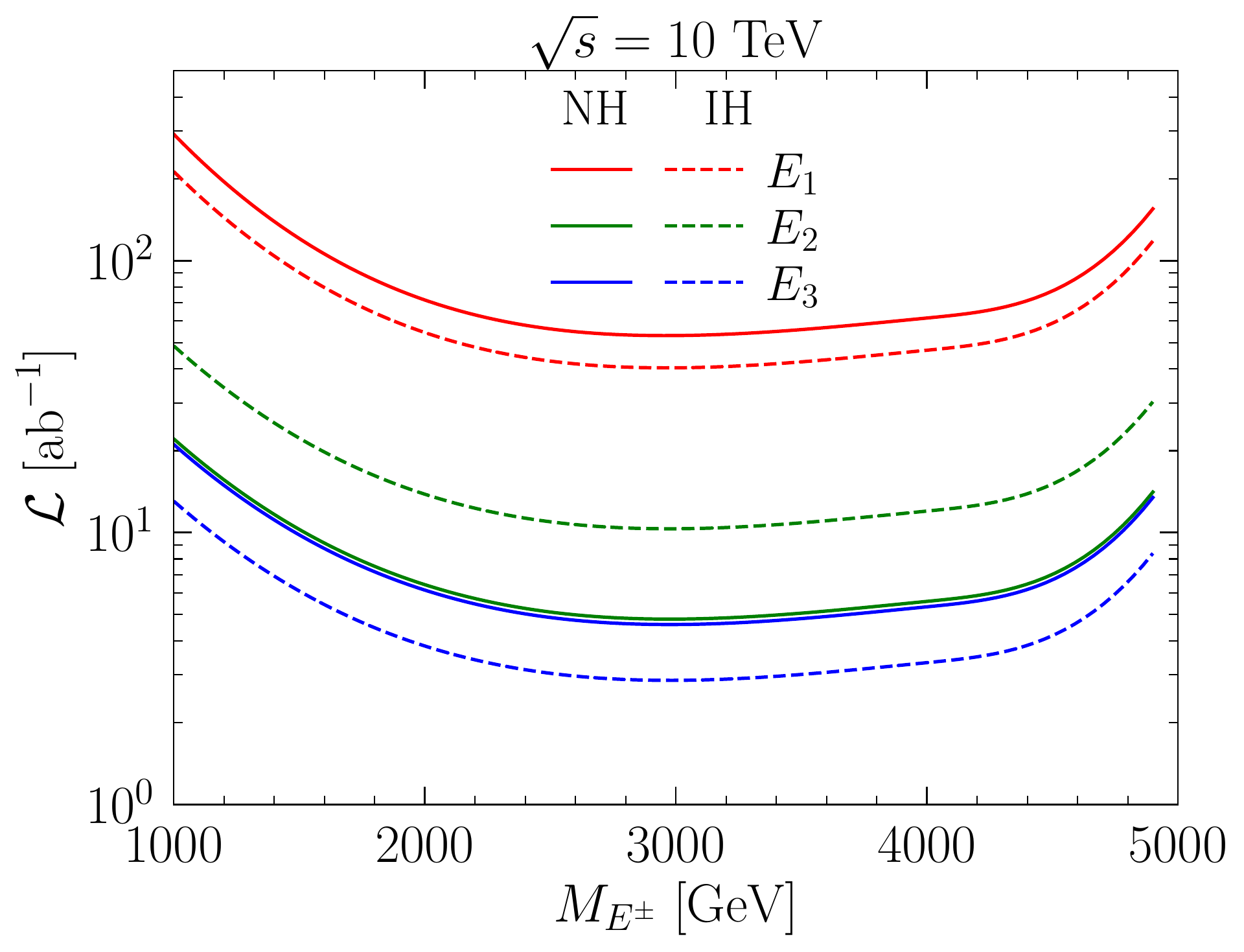}
\end{center}
\caption{The integrated luminosities for 5$\sigma$ significance versus $M_E$ for $E^+E^-\to ZZ\mu^+\mu^-$ in muon colliders with $\sqrt{s}=3$ TeV (left) and 10 TeV (right), for $E_{1,2,3}$ and NH (solid lines) or IH (dashed lines).
}
\label{diagram:Luminosity of E to Zmu}
\end{figure}

\begin{figure}[h!]
\begin{center}
\includegraphics[width=0.46\textwidth]{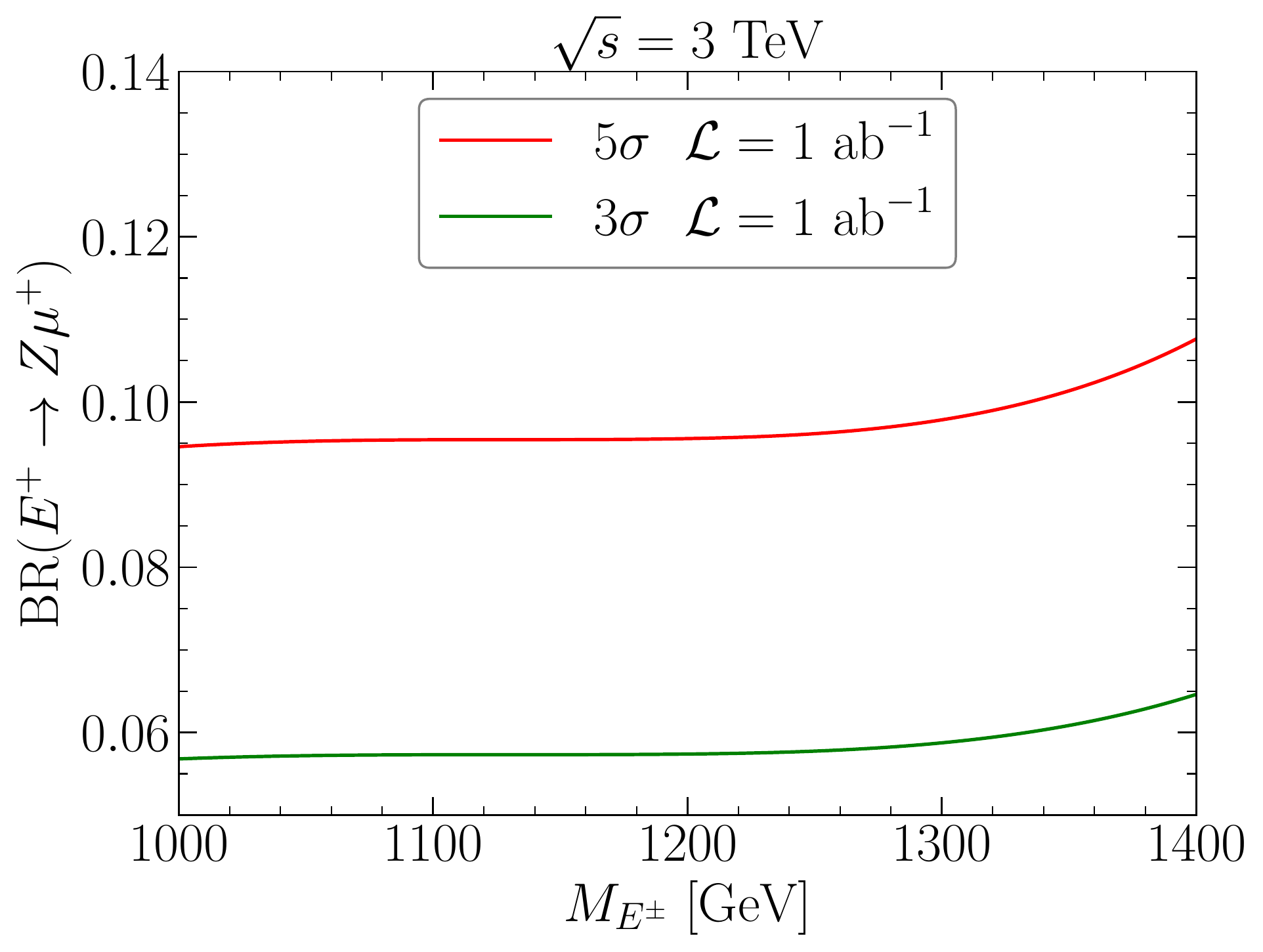}
\includegraphics[width=0.46\textwidth]{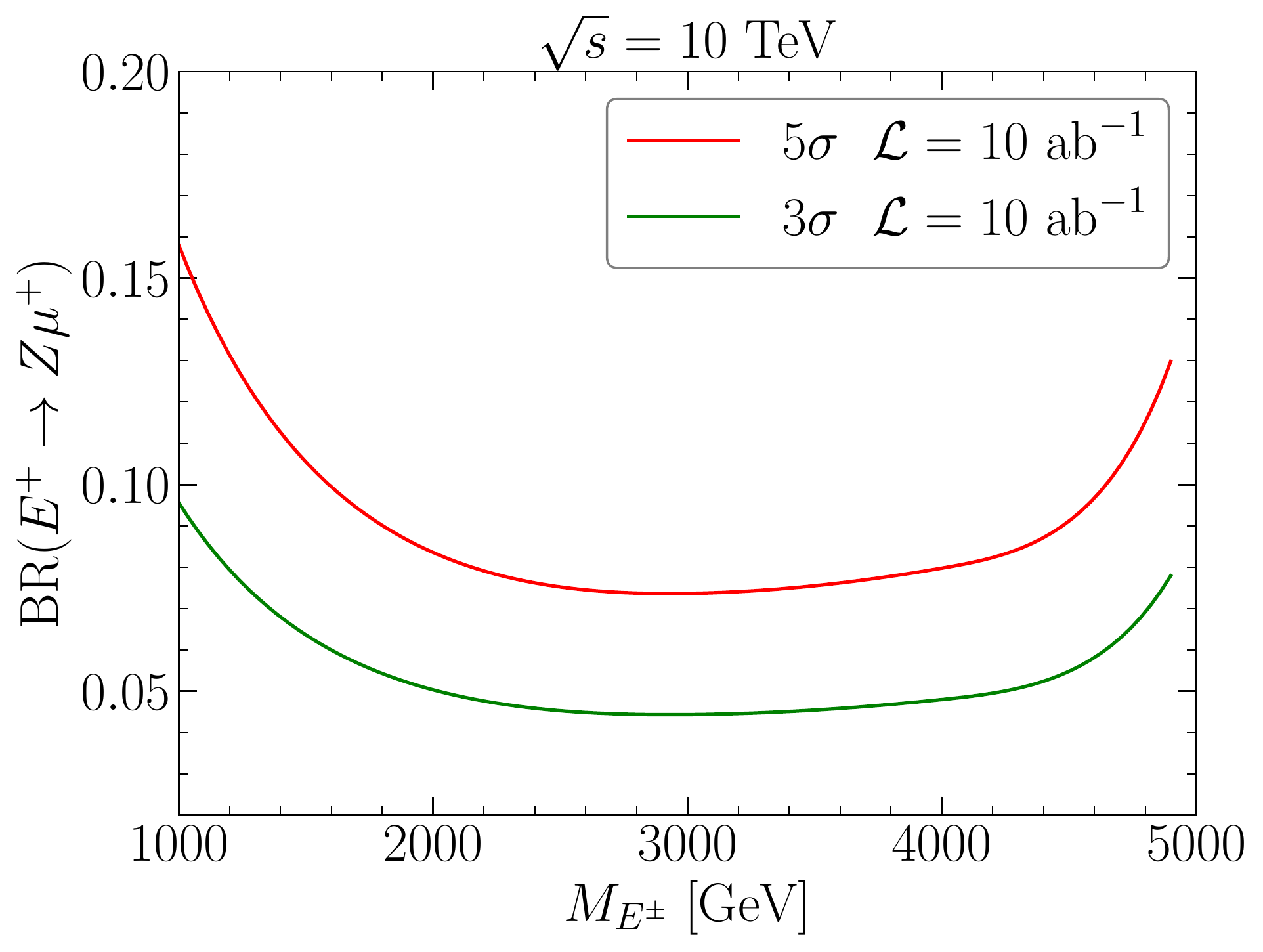}
\end{center}
\caption{The reachable branching ratio of $E^\pm\to Z\mu^\pm$ corresponding to 3$\sigma$ and 5$\sigma$ significance versus $M_E$ for $E^+E^- \to ZZ\mu^+\mu^-$ in muon colliders with $\sqrt{s}=3$ TeV, $\mathcal{L}=1~{\rm ab}^{-1} $ (left) and $\sqrt{s}=10$ TeV, $\mathcal{L}=10~{\rm ab}^{-1}$ (right).
}
\label{diagram:BR of E to Zmu}
\end{figure}

For the local significance analysis, we apply the benchmark decay branching fraction of $E^\pm\to \mu^\pm Z$ for $N_{1,2,3}$ in both NH and IH mass patterns in Table~\ref{BRN}.
As a result, in Fig.~\ref{diagram:Luminosity of E to Zmu}, we show the integrated luminosities for $5\sigma$ significance in the muon collider with $\sqrt{s}=3$ TeV (left) and $\sqrt{s}=10$ TeV (right). With the optimistic integrated luminosity of 1 ab$^{-1}$ for $\sqrt{s}=3$ TeV or 10 ab$^{-1}$ for $\sqrt{s}=10$ TeV, both $N_2$ and $N_3$ can be discovered. One can distinguish $E_2$ and $E_3$ for IH and they are not distinguishable for NH. To discover $N_1$, we need much higher luminosity.
A 5$\sigma$ significance for $\sqrt{s}=3$ TeV will be reached with the integrated luminosity above 1 ab$^{-1}$ for NH and IH.
The reachable branching ratio of $E^\pm\to Z\mu^\pm$ corresponding to 3$\sigma$ and 5$\sigma$ significance is given in Fig.~\ref{diagram:BR of E to Zmu}, with $\sqrt{s}=3$ TeV, $\mathcal{L}=1~{\rm ab}^{-1}$ (left) and $\sqrt{s}=10$ TeV, $\mathcal{L}=10~{\rm ab}^{-1}$ (right).
One can see that BR$(E^\pm\to Z\mu^\pm)$ below 10.7\% (6.4\%) can be reached for 5$\sigma$ (3$\sigma$) significance for $\sqrt{s}=3$ TeV and $\mathcal{L}=1~{\rm ab}^{-1}$. For $\sqrt{s}=10$ TeV and $\mathcal{L}=10~{\rm ab}^{-1}$, the branching fraction can be as large as 13.0\% (7.8\%) for 5$\sigma$ (3$\sigma$) significance.

%%%%%%%%%%%%%%%%%%%%%%%%%%%%%%%%%%%%%%
\subsection{$E^{\pm} N$ Associated Production}
\label{sec:EN}
%%%%%%%%%%%%%%%%%%%%%%%%%%%%%%%%%%%%%%

\begin{figure}[htbp]
\makebox[\linewidth][c]{%
\centering
\includegraphics[width=0.5\textwidth]{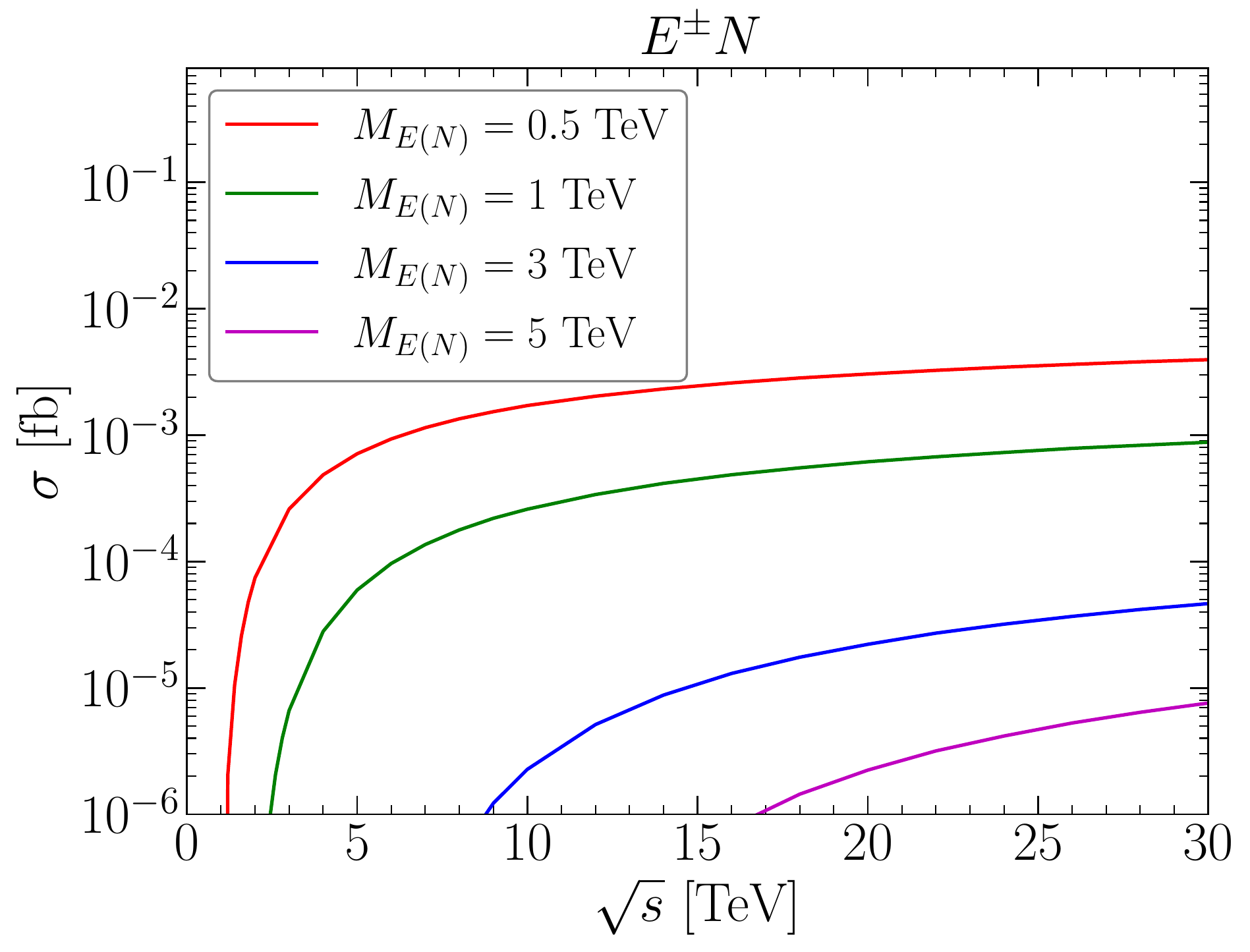}
\includegraphics[width=0.5\textwidth]{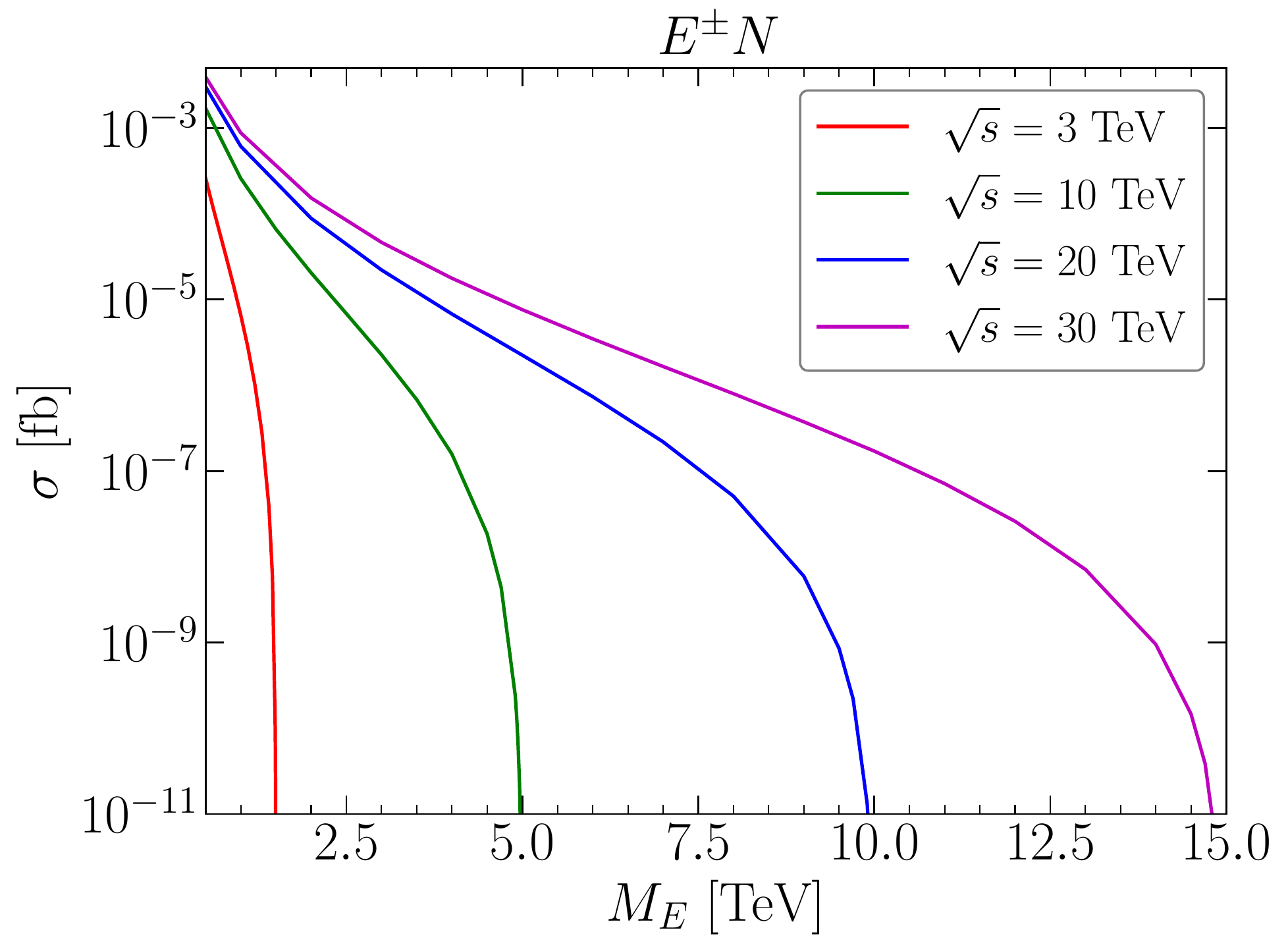}
}
\caption{Cross sections of $E^{\pm} N$ associated production as a function of $\sqrt{s}$ (left) and $M_{E,N}$ (right) at muon colliders by VBF process.
}
\label{fig:EN_production_sqrts}
\end{figure}

The $E^{\pm}N$ associated production is only from VBF processes initiated by $\gamma / Z $ and $W^{\pm}$. The cross sections are shown in Fig.~\ref{fig:EN_production_sqrts}, as a function of $\sqrt{s}$ (left) and $M_{E(N)}$ (right), and are enhanced by EW PDF at higher beam energies. We find that muon collider with energy $3$~TeV or 10~TeV can not generate enough signal events, and we thus take $\sqrt{s}=30$ TeV in the following studies. For $\sqrt{s}=30$ TeV, under the assumption that the masses of charged and neutral triplet leptons are degenerate, the cross sections are less than $10^{-3}$ fb for triplet lepton masses greater than 1 TeV. The associated production of charged and neutral heavy leptons followed by decay modes $E^\pm\to \ell^\pm Z$ and $N\to \ell^\pm W^\mp$ can lead to the rare signature of lepton number violation with same-sign charged leptons
\begin{eqnarray}
E^{\pm}N \to Z W^{\mp} \mu^{\pm} \mu^{\pm}\;.
\end{eqnarray}
We will only consider the sensitivity of muon collider with $\sqrt{s}=30$ TeV to the distinctive LNV signature.

\begin{figure}[htbp]
\makebox[\linewidth][c]{%
\centering
\includegraphics[width=0.5\textwidth]{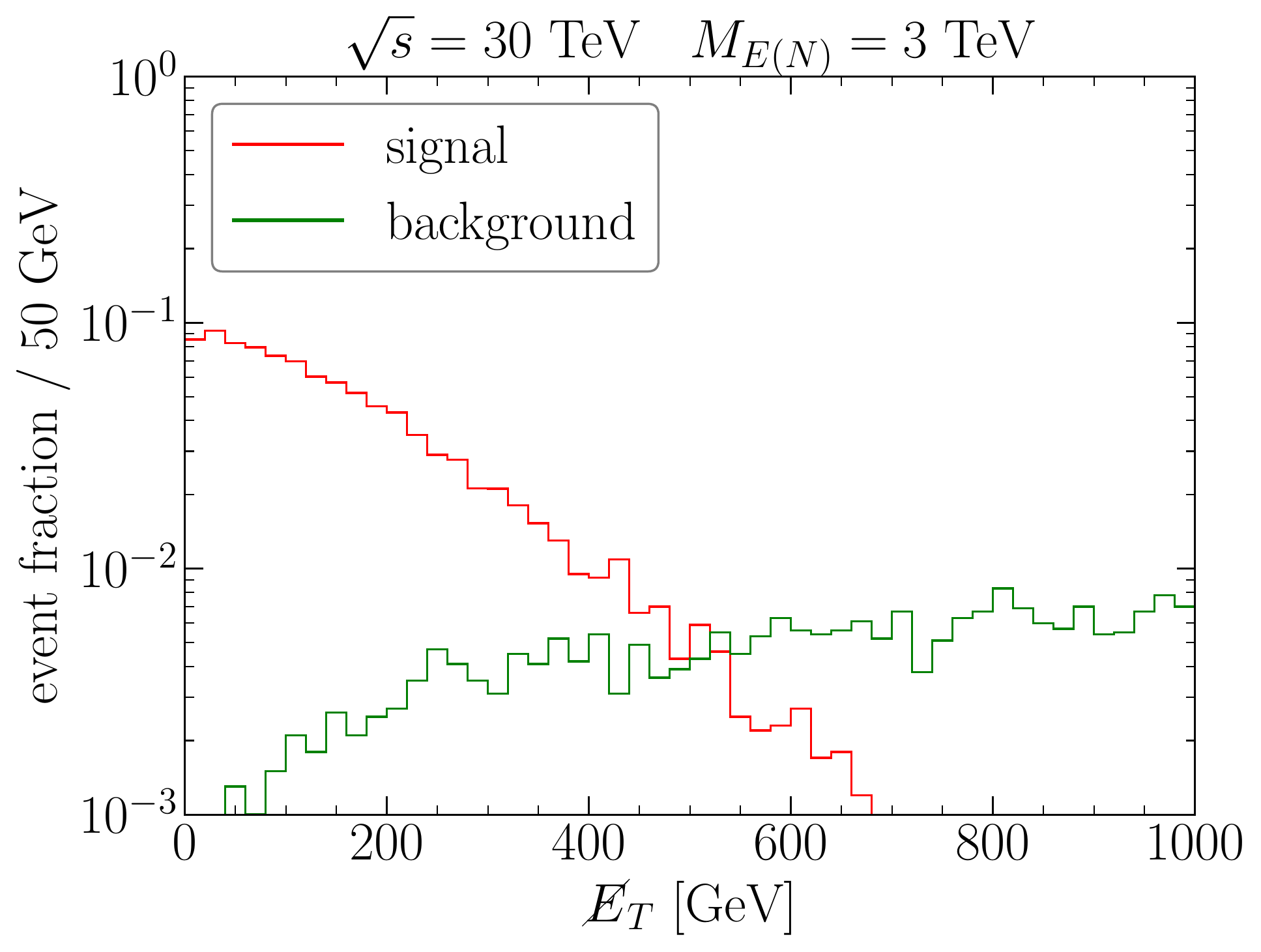}
\includegraphics[width=0.5\textwidth]{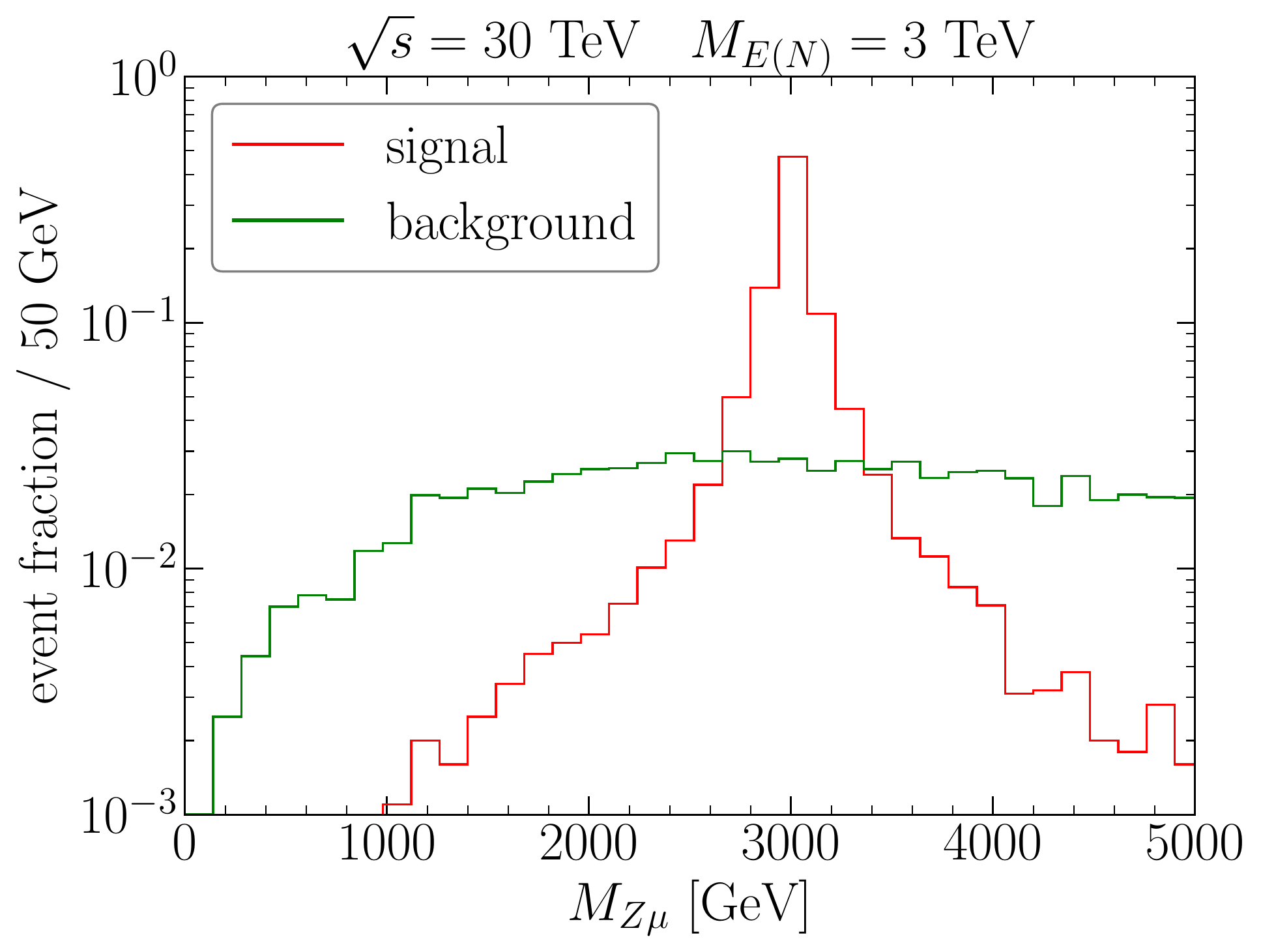}
}
\caption{
The normalized $\cancel{E}_T$ (left) and invariant mass $M_{Z\mu}$ (right) distributions of the heavy leptons in the signal $E^{\pm}N \to Z W^{\mp} \mu^{\pm}  \mu^{\pm}$ and the background for $\sqrt{s}= 30$ TeV and $M_{E(N)} = 3$ TeV.
}
\label{fig:SB-mmwz-30TeV}
\end{figure}

The leading SM background to this channel becomes
\begin{eqnarray}
Z W^{\mp} W^{\pm} W^{\pm}\;,
\end{eqnarray}
followed by decay $W^{\pm} \to \mu^{\pm} \nu_{\mu}(\bar{\nu}_{\mu})$. We simulate VBF production of the above background using MadGraph at the parton level.
Firstly, we employ the basic cuts for the gauge bosons in final states
\begin{align}
&p_{T}(W,Z)>300~{\rm GeV}\, ,~~\left| \eta(W, Z) \right|< 3.0\;.
\end{align}
We also employ further judicious cuts to reduce the background
\begin{itemize}
\item
For a few TeV $E^{\pm}$ and $N$, we set the advanced $p_T$ cuts on the gauge bosons as follows
\begin{eqnarray}
p_{T}(W)> \frac{M_N}{3} \, , \quad p_{T}(Z)> \frac{M_E}{3}\;.
\end{eqnarray}
\item
Note that the $\nu_{\mu}$ or $\bar{\nu}_{\mu}$ from $W$ boson decay only appear in the SM background.
For our signal, the $E^\pm N$ or $NN$ production is induced by the VBF process in which there is no missing neutrino. Nevertheless, in the process of event analysis, we consider the smearing of the transverse momentum with a resolution of the detector mentioned below Eq.~(\ref{eq:para}). As a result, there appears a small transverse missing energy in the signal. As shown in left panel of Fig.~\ref{fig:SB-mmwz-30TeV}, we veto the events with large missing energy from neutrinos
\begin{eqnarray}
\cancel{E}_{T} < 500~{\rm GeV}\;.
\end{eqnarray}
\item To suppress the continuum background, we reconstruct the heavy lepton $E^{\pm}$ and $N$. Firstly, we pair the final particles $W \mu$ and $Z \mu$ according to the feature that the real combination comes from heavy triplets with degenerate masses. The reconstructed heavy lepton masses are shown in the right panel of Fig.~\ref{fig:SB-mmwz-30TeV}. Then we take the mass windows of the resonances
\begin{eqnarray}
\left | M_{Z\mu_1}-M_E \right|< \frac{M_E}{5} \, , \quad \left | M_{W\mu_2}-M_N \right| < \frac{M_N}{5}\;.
\end{eqnarray}
\end{itemize}

\begin{figure}[htbp]
\makebox[\linewidth][c]{%
\centering
\includegraphics[width=0.5\textwidth]{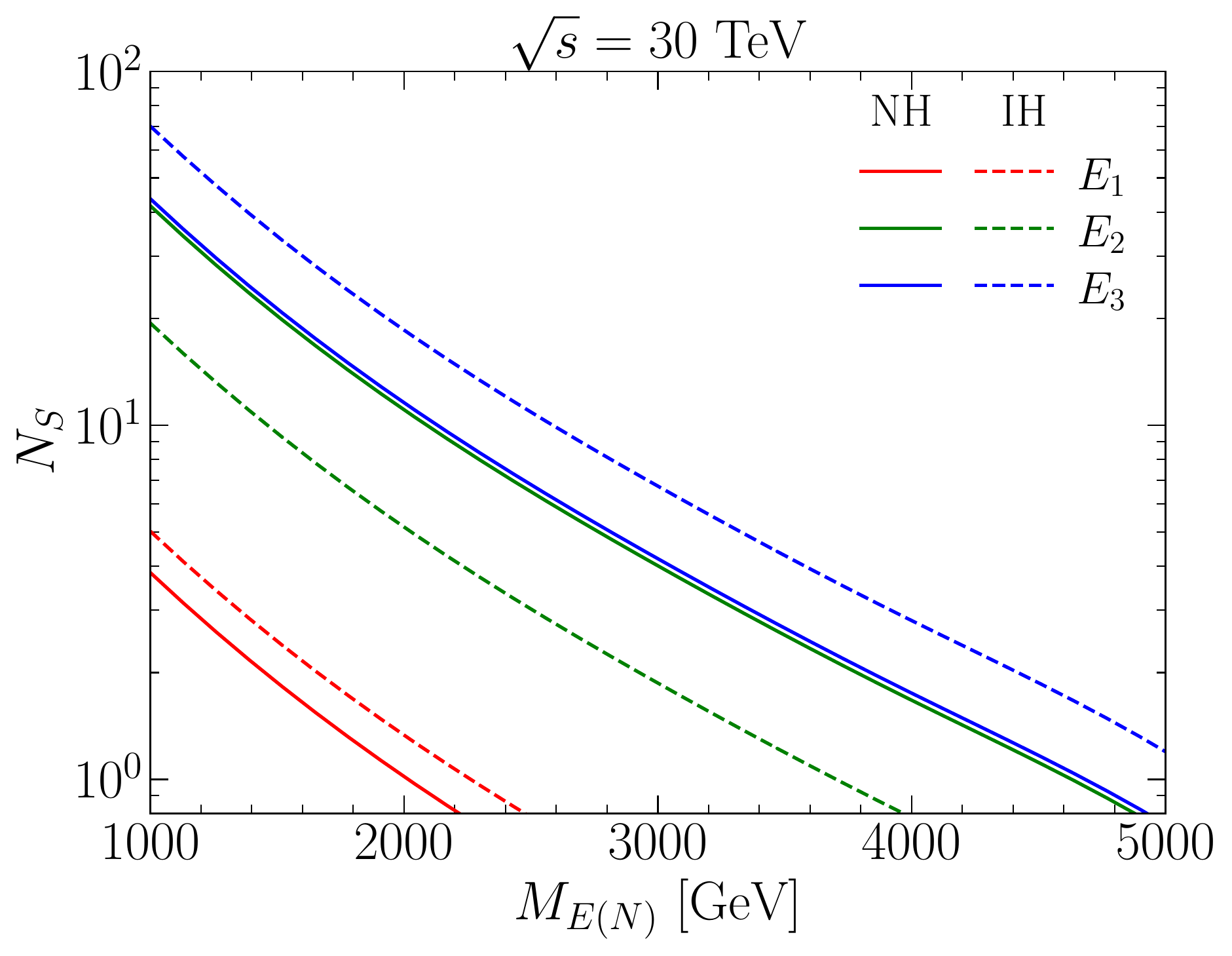}
\includegraphics[width=0.5\textwidth]{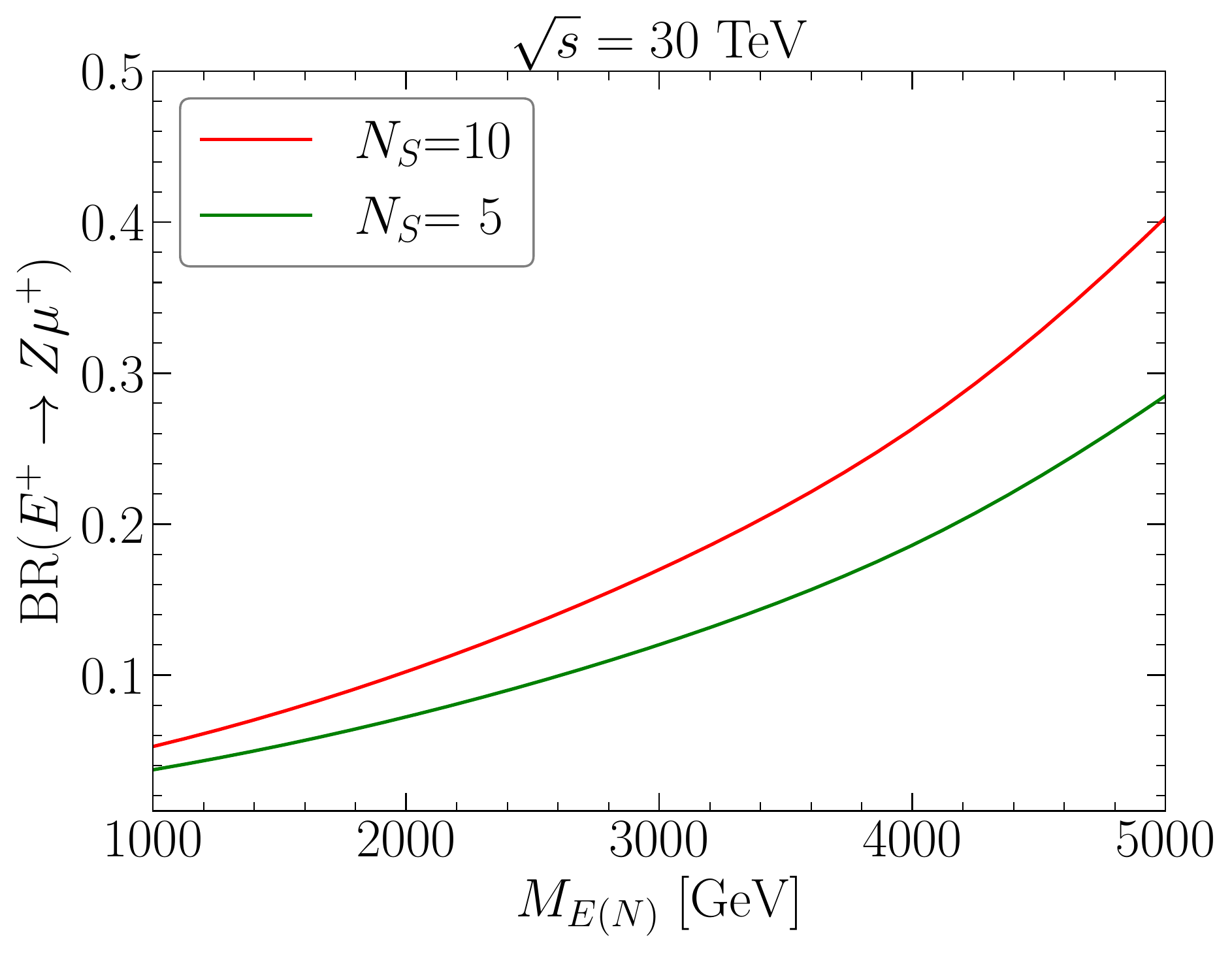}
}
\caption{
Left: The number of signal events versus $M_{E(N)}$ for $E^\pm N \to W^\mp Z \mu^\pm\mu^\pm$ at muon collider with $\sqrt{s}=30$ TeV and $\mathcal{L}=90~{\rm ab}^{-1}$, for $E(N)_{1,2,3}$ and NH (solid lines) or IH (dashed lines). Right: The reachable branching ratio of $E^\pm\to Z\mu^\pm$ or $N\to W^\mp \mu^\pm$ corresponding to $N_S=5$ and $N_S=10$ versus $M_{E(N)}$ for $E^\pm N \to W^\mp Z \mu^\pm\mu^\pm$ at muon collider with $\sqrt{s}=30$ TeV and $\mathcal{L}=90~{\rm ab}^{-1}$.
}
\label{fig:S-mmwz-30TeV}
\end{figure}

One can efficiently suppress the SM background and the cut efficiency becomes $\epsilon_B \sim 0.1\%$ by using the above cuts, so it is enough to only consider the signal events. The event number of signal is given by

\begin{align}
N_S=&\sigma_{S}^{\rm{VBF}} \times \epsilon_{S}^{\rm{VBF}}\times\epsilon^{\rm VBF}_{S,W/Z} \times {\rm BR}(E^{\pm}\to Z \mu^{\pm})\times {\rm BR}(N \to W^{\mp}\mu^{\pm})\\ \nonumber
&\times \mathcal{L} \times {\rm BR}(W^{\pm} \to q\bar{q}')\times {\rm BR}(Z \to q\bar{q})  \;,\nonumber
\end{align}
where $\epsilon^{\rm VBF}_{S,W/Z}$ denotes the efficiency of reconstructing $W$ and $Z$ bosons, ${\rm BR}(E^{\pm}\to Z \mu^{\pm})\approx {\rm BR}(N \to W^{\mp}\mu^{\pm})$ as shown in Table~\ref{BRN}.
The number of signal events at muon collider with $\sqrt{s}=30$ TeV is shown in left panel of Fig.~\ref{fig:S-mmwz-30TeV}, with the integrated luminosity of 90 ${\rm ab}^{-1}$. One can see that the three heavy triplet leptons and two neutrino mass patterns exhibit different event numbers. For $M_{E(N)}$ greater than 6 TeV, it is impossible to receive more than one signal event. The reachable branching ratio of $E^{\pm}\to Z \mu^{\pm}$ and $N \to W^{\mp}\mu^{\pm}$ corresponding to $N_S=5$ and $N_S=10$ is given in the right panel of Fig.~\ref{fig:S-mmwz-30TeV}. ${\rm BR}(E^{\pm}\to Z \mu^{\pm})$ below 40.3$\%$ (28.5$\%$) can be reached for $N_S=10$ ($N_S=5$) and $M_{E(N)}<5$ TeV at muon collider with $\sqrt{s}=30$ TeV and $\mathcal{L}=90$ ${\rm ab}^{-1}$.

%%%%%%%%%%%%%%%%%%%%%%%%%%%%%%%%%%%%%%
\subsection{$NN$ Pair Production}
%%%%%%%%%%%%%%%%%%%%%%%%%%%%%%%%%%%%%%

\begin{figure}[htbp]
\makebox[\linewidth][c]{%
\centering
\includegraphics[width=0.5\textwidth]{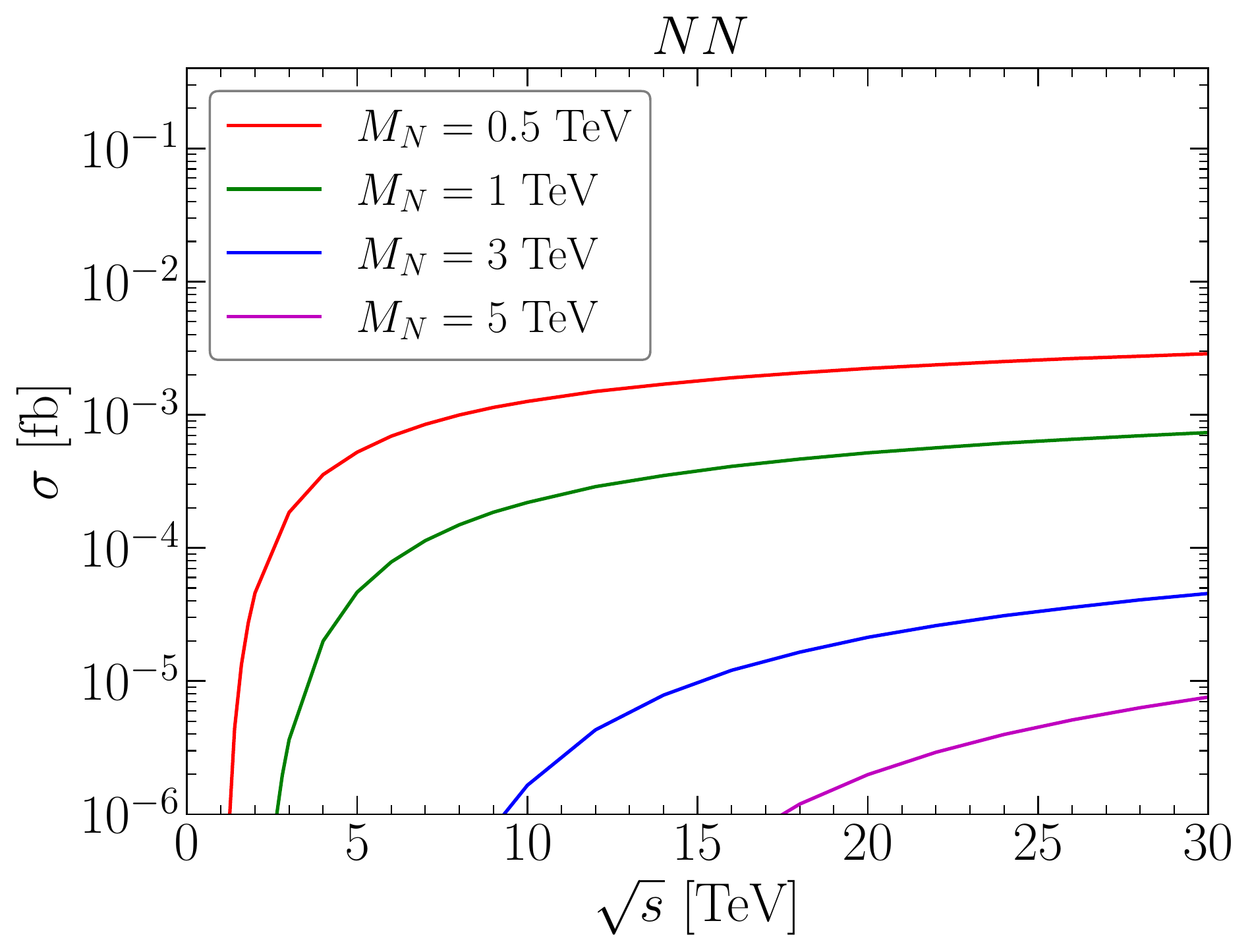}
\includegraphics[width=0.5\textwidth]{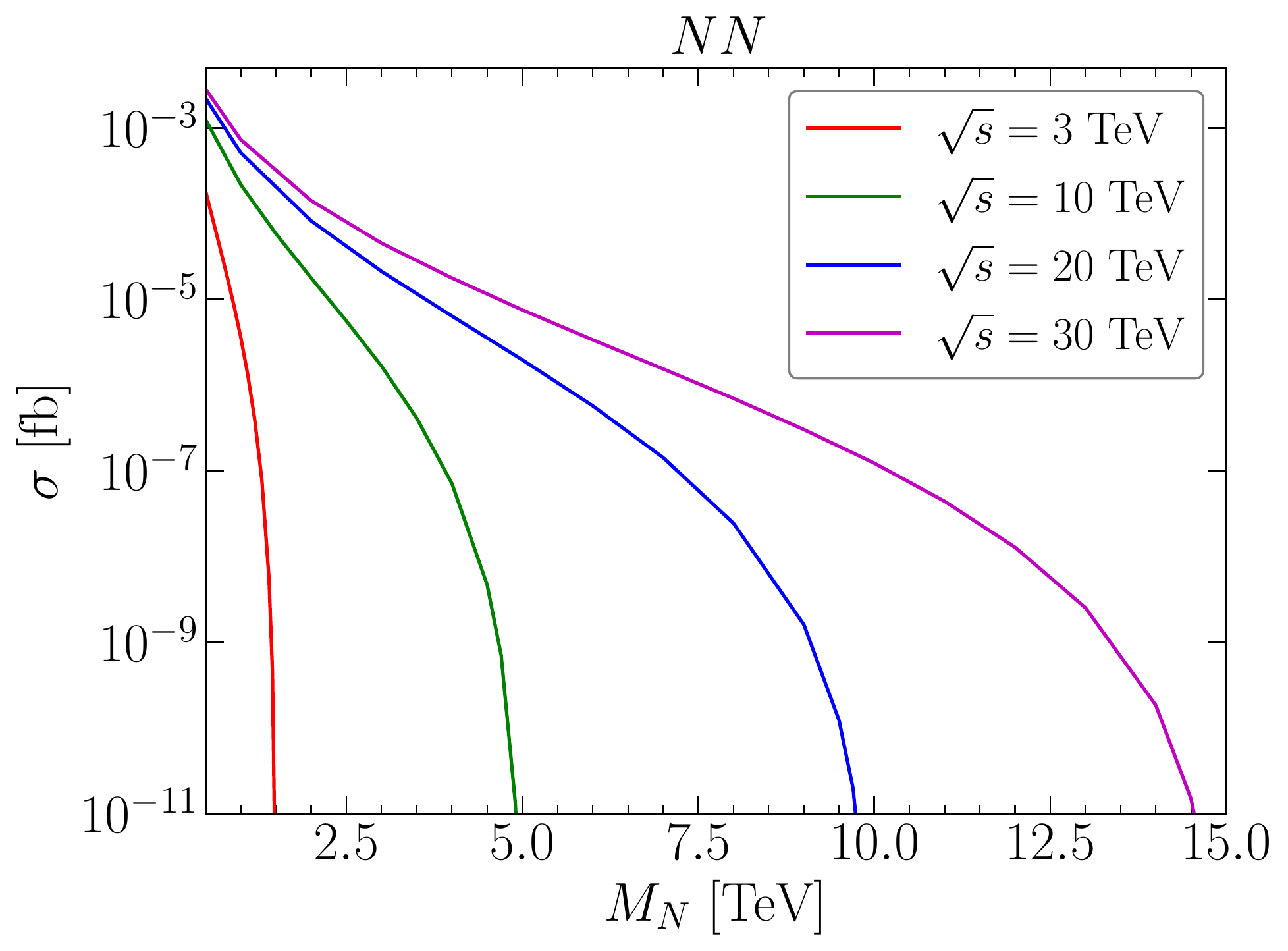}
}
\caption{Cross sections of $NN$ pair production as a function of $\sqrt{s}$ (left) and $M_{N}$ (right) at muon colliders by VBF process.
}
\label{fig:NN_production_sqrts}
\end{figure}

\begin{figure}[htbp]
\makebox[\linewidth][c]{%
\centering
\includegraphics[width=0.5\textwidth]{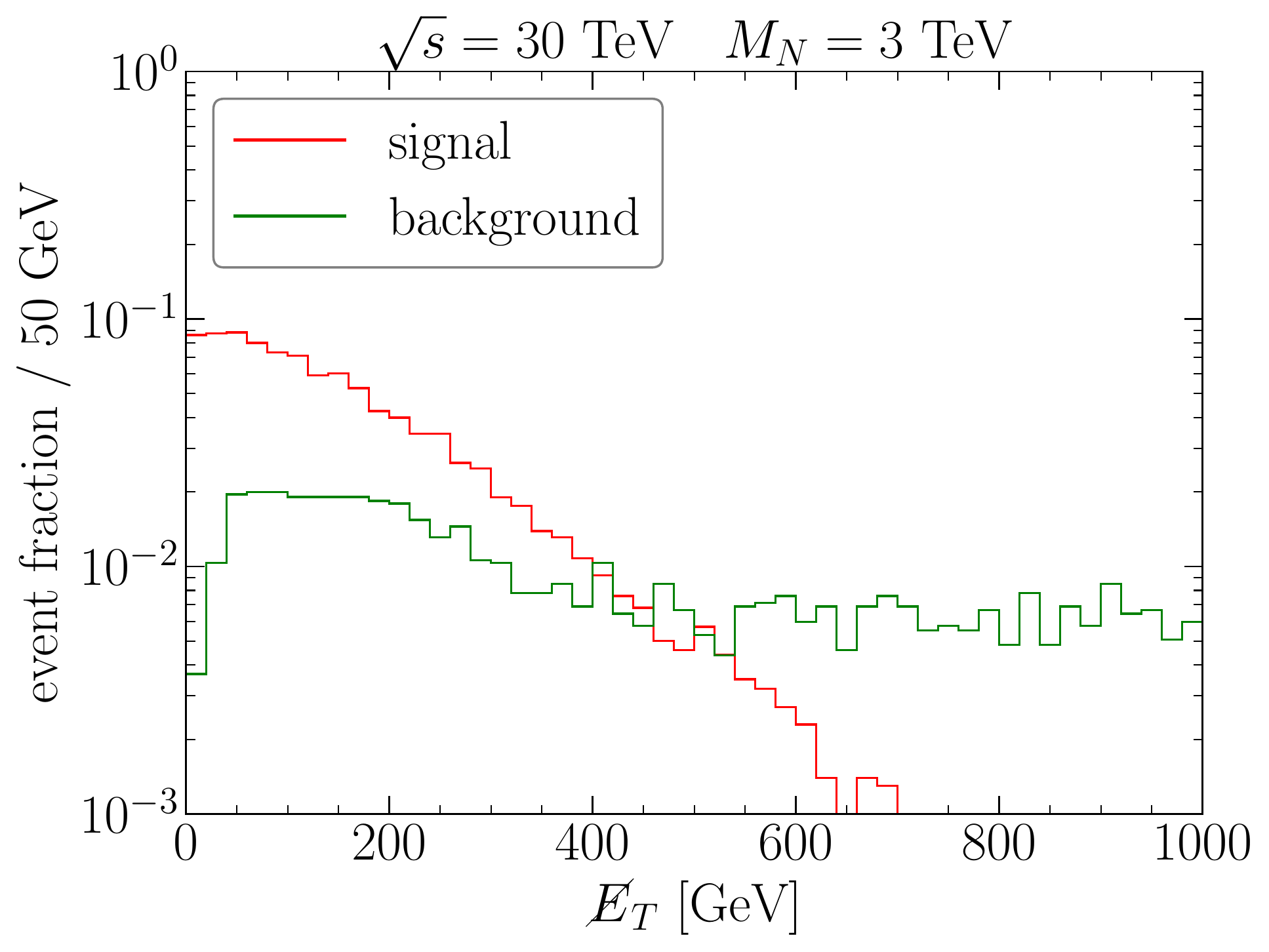}
\includegraphics[width=0.5\textwidth]{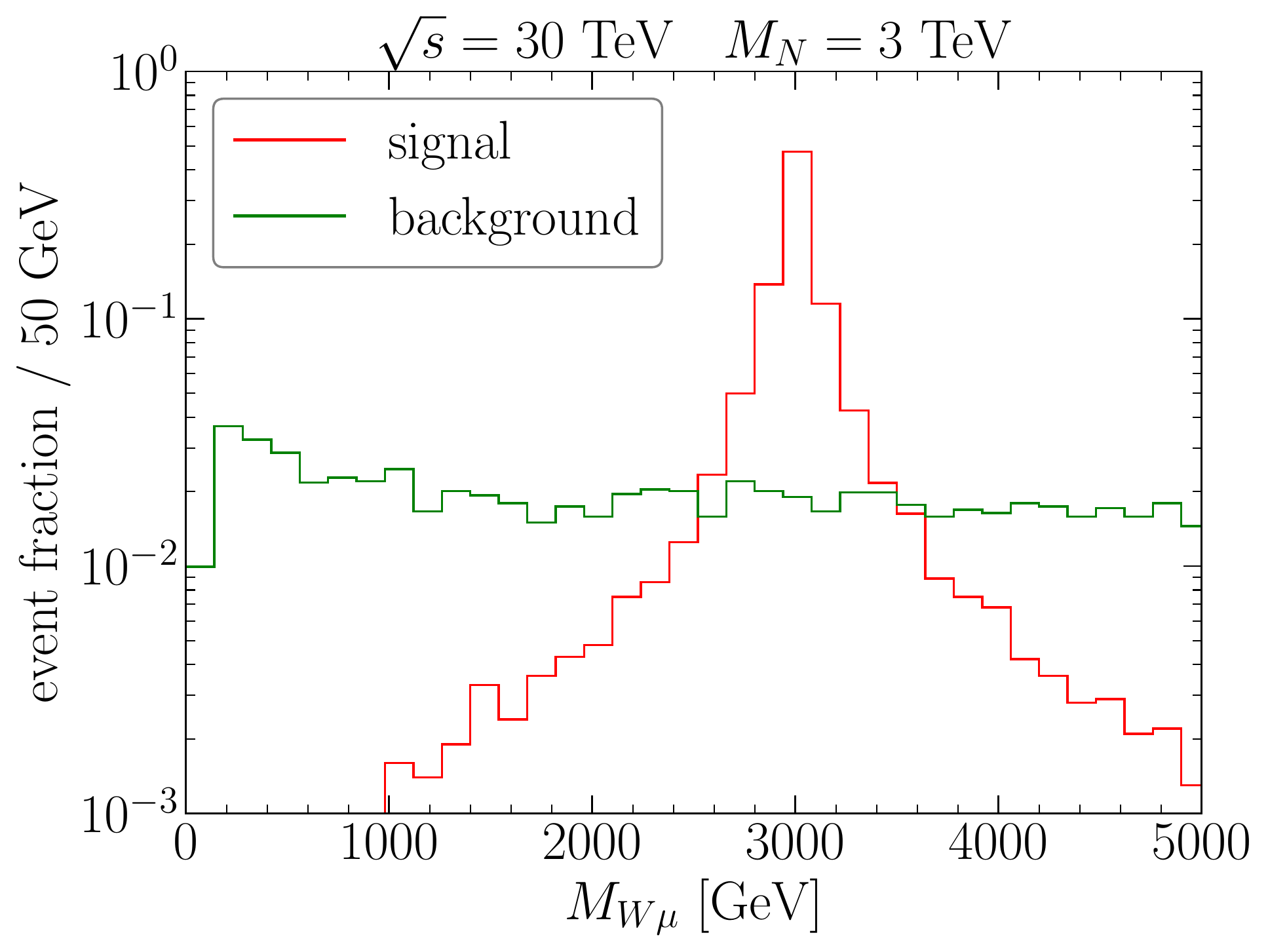}
}
\caption{
The normalized $\cancel{E}_T$ (left) and invariant mass $M_{W\mu}$ (right) distributions of the heavy leptons in the signal $ N N \to  W^{\mp} W^{\mp} \mu^{\pm}  \mu^{\pm}$ and the background for $\sqrt{s}= 30$ TeV and $M_N = 3$ TeV.
}
\label{fig:SB-mmww-30TeV}
\end{figure}

In this subsection, we discuss the $NN$ pair production from VBF process initiated by $W^+W^-$. The VBF cross sections as a function of $\sqrt{s}$ and $M_N$ are shown in Fig.~\ref{fig:NN_production_sqrts}. The cross sections are slightly smaller than those of $E^{\pm}N$ associated production as there is only $W^+W^-$ initiated VBF process. We will only consider the distinctive LNV signature at muon collider with $\sqrt{s}=30$ TeV, i.e.,
\begin{eqnarray}
N N \to W^{\mp} W^{\mp} \mu^{\pm} \mu^{\pm}\;.
\end{eqnarray}
The leading SM background becomes
\begin{eqnarray}
W^+W^+W^-W^-\;,
\end{eqnarray}
followed by decay $W^{\pm} \to \mu^{\pm}\nu_{\mu}(\bar{\nu}_{\mu})$ for one pair of same-sign $W$ bosons. The judicious cuts to reduce the background are
\begin{itemize}
\item
For a few TeV $N$, we set the advanced $p_T$ cuts for the leading $W$ boson and $\mu^{\pm}$ as follows
\begin{eqnarray}
p_{T}^{\rm max}(W)> \frac{M_N}{3} \, , \quad p_{T}^{\rm max}(\mu)> \frac{M_N}{3}\;.
\end{eqnarray}
\item
Since the $\nu_{\mu}$ or $\bar{\nu}_{\mu}$ only appears in the SM background, we can set the cuts on missing energy from neutrinos as shown in left panel of Fig.~\ref{fig:SB-mmww-30TeV}
\begin{eqnarray}
\cancel{E}_{T} < 500~{\rm GeV}\;.
\end{eqnarray}
\item
In order to reconstruct the heavy neutrino $N$, we pair the particles $W \mu$ according to the feature that the real combination comes from heavy triplets with degenerate masses. The invariant mass distribution is shown in the right panel of Fig.~\ref{fig:SB-mmww-30TeV}. Then we take the mass windows of the resonances
\begin{eqnarray}
\left | M_{W\mu}-M_N \right| < \frac{M_N}{5}\;.
\end{eqnarray}
\end{itemize}

\begin{figure}[htbp]
\makebox[\linewidth][c]{%
\centering
\includegraphics[width=0.5\textwidth]{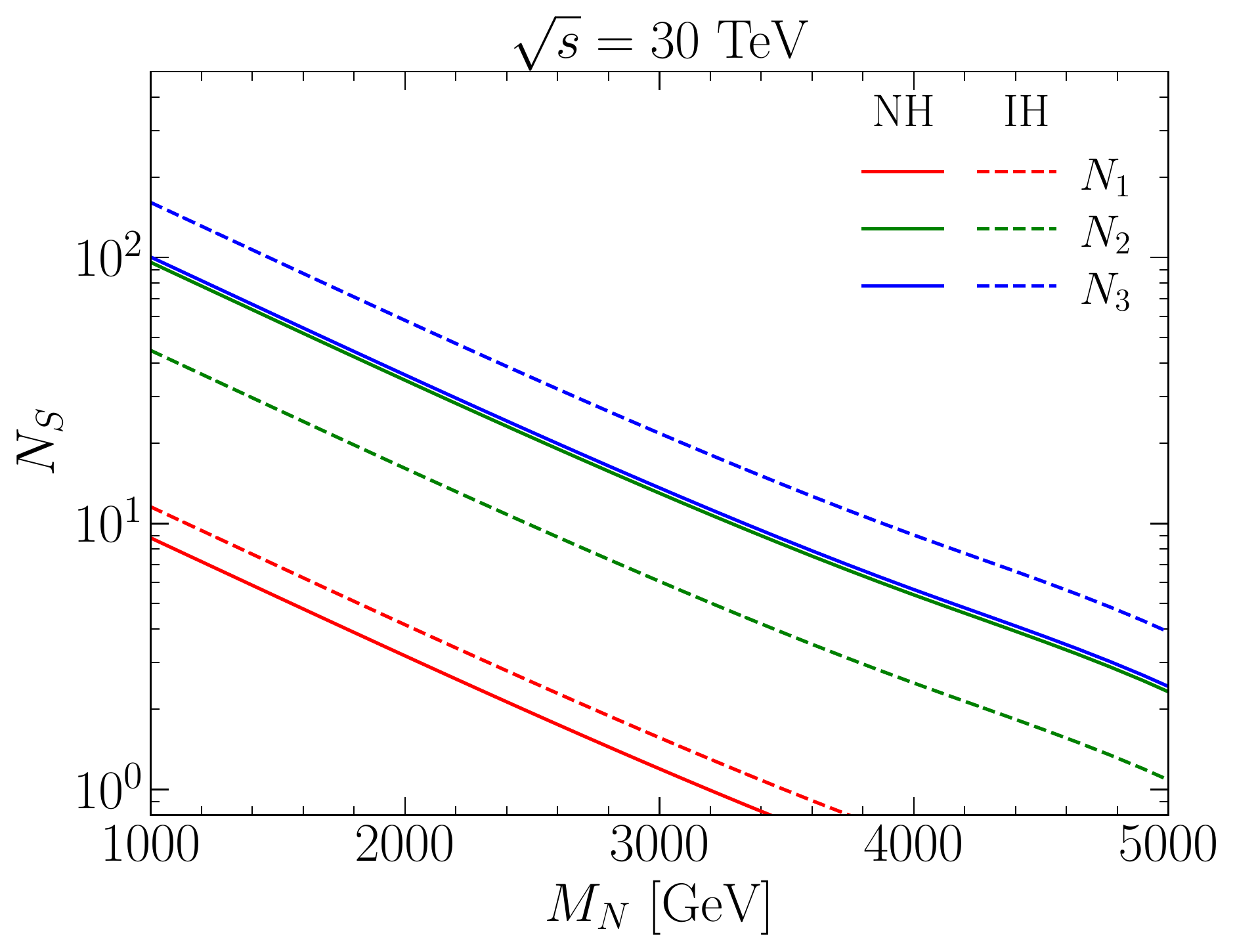}
\includegraphics[width=0.5\textwidth]{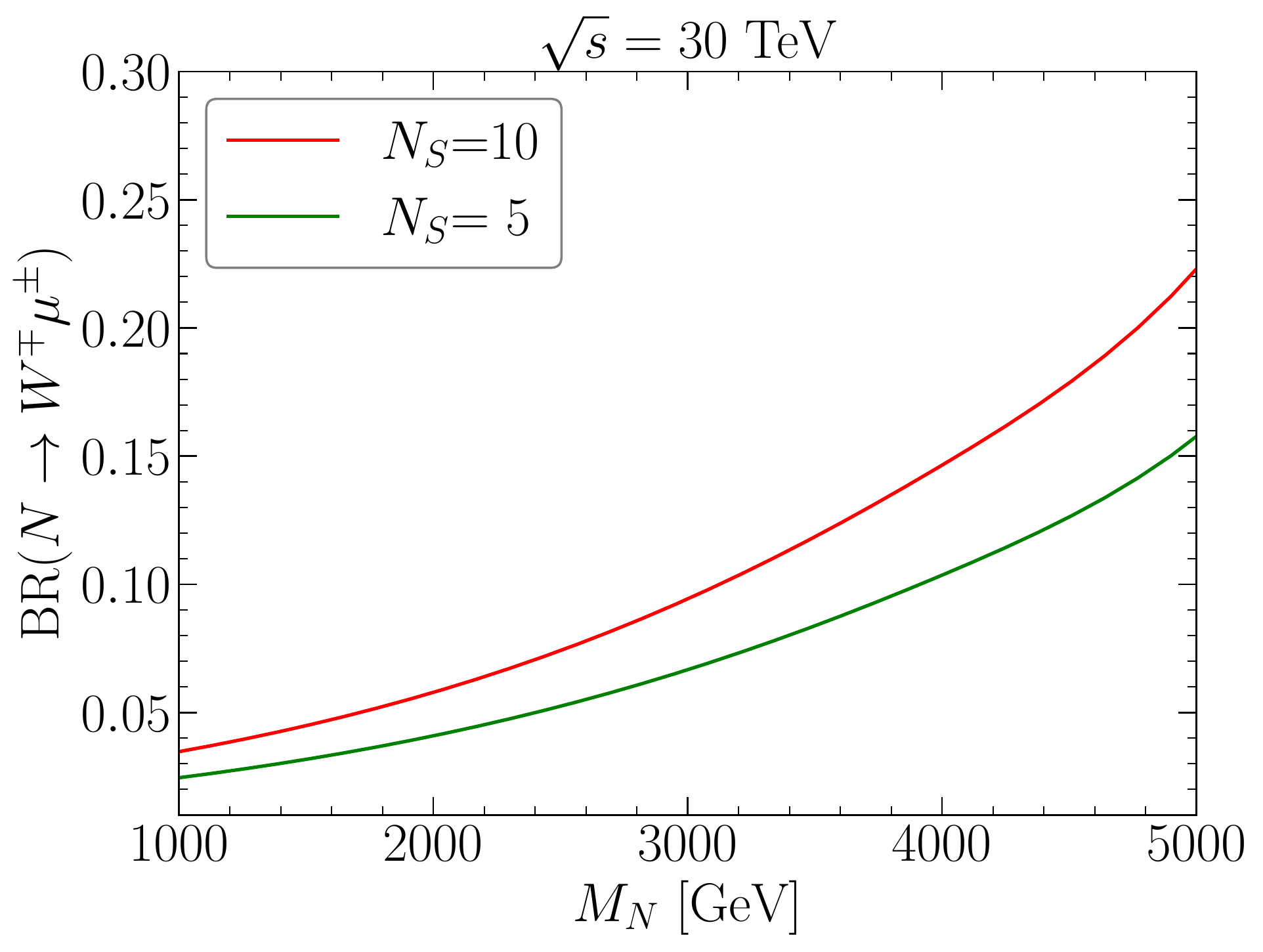}
}
\caption{
Left: The number of signal events versus $M_N$ for $ N N \to  W^{\mp} W^{\mp} \mu^{\pm}  \mu^{\pm} $ at muon collider with $\sqrt{s}=30$ TeV and $\mathcal{L}=90~{\rm ab}^{-1}$, for $N_{1,2,3}$ and NH (solid lines) or IH (dashed lines). Right: The reachable branching ratio of $N\to W^\mp \mu^\pm$ corresponding to $N_S=5$ and $N_S=10$ versus $M_N$ for $ N N \to  W^{\mp} W^{\mp} \mu^{\pm}  \mu^{\pm} $ at muon collider with $\sqrt{s}=30$ TeV and $\mathcal{L}=90~{\rm ab}^{-1}$.
}
\label{fig:S-mmww-30TeV}
\end{figure}

Since the SM background is extremely suppressed, we will only consider the signal events. The event number of signal is given by
\begin{eqnarray}
N_S=\sigma_{S}^{\rm{VBF}} \times \epsilon_{S}^{\rm{VBF}}\times\epsilon^{\rm VBF}_{S,W}\times  {\rm BR}^2(N \to W^{\mp}\mu^{\pm}) \times \mathcal{L}\times {\rm BR}(W^{\pm} \to q\bar{q}')^2 \times 2\ \;,
\end{eqnarray}
where the factor 2 accounts for the charge conjugation of decay products, as shown in the left panel of Fig.~\ref{fig:S-mmww-30TeV}.
For $M_N\gtrsim 6$ TeV, it is almost impossible to obtain more than one signal event. The reachable branching ratio of $N \to W^{\mp} \mu^{\pm} $ is given in the right panel of Fig.~\ref{fig:S-mmww-30TeV}. ${\rm BR}(N \to W^{\mp} \mu^{\pm})$ below $22.2\%~(15.8\%)$ can be reached for $N_S=10~(N_S=5)$ and $M_N<5$ TeV at muon collider with $\sqrt{s}=30$ TeV and $\mathcal{L}=90~{\rm ab}^{-1}$.

\begin{figure}[htbp]
\centering
\includegraphics[width=0.49\textwidth]{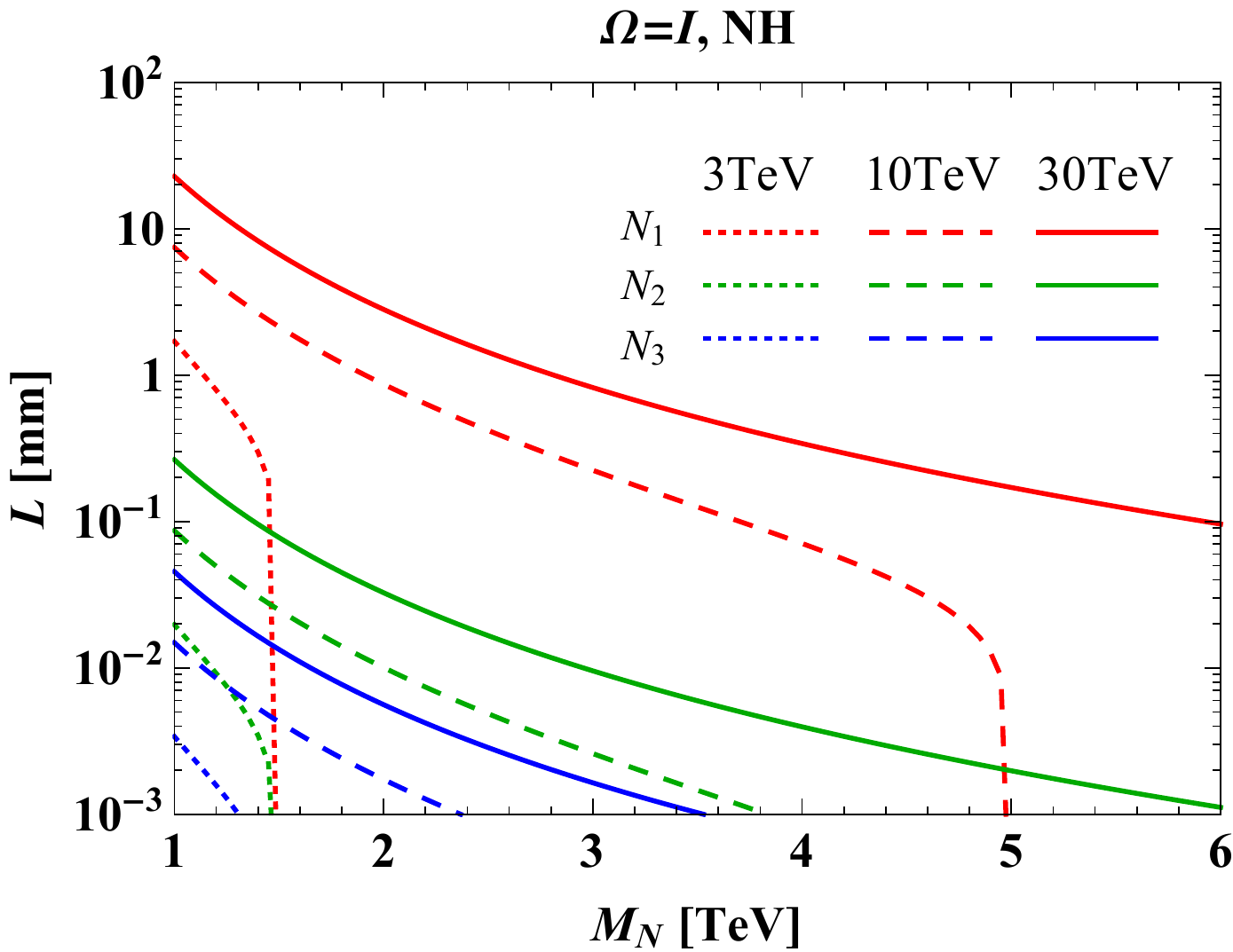}
\includegraphics[width=0.49\textwidth]{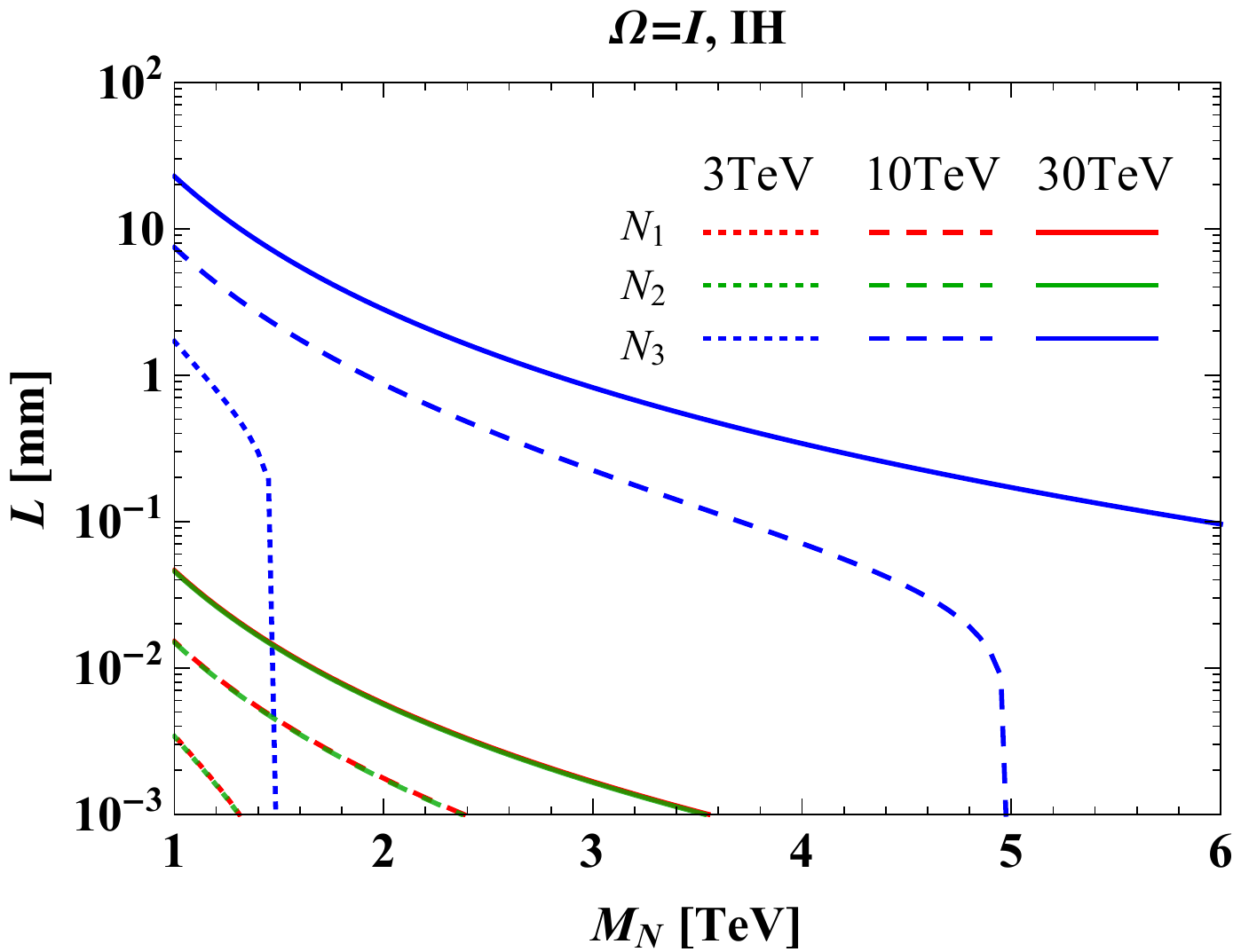}
\caption{Decay length versus $M_N$ for different heavy neutrinos $N_i~(i=1,2,3)$ and different neutrino mass hierarchies NH~(left) and IH~(right) at muon collider with energy $3$~TeV (dotted line), $10$~TeV (dashed line) and $30$~TeV (solid line) respectively.
}
\label{fig:lifetime}
\end{figure}

Finally, we comment on the decay length and possible displaced vertex of heavy triplet leptons. The total decay widths of heavy triplet leptons are proportional to $|V_{\ell N}|^2\sim m_\nu/M_\Sigma$ which may lead to long decay length.
We calculate the decay length of the triplet leptons by the formula $L= \gamma\beta c\tau$, where $\tau=1/\Gamma$ is the life-time, $\gamma$ is the boost factor and $\beta$ is the ratio of velocity $v_\Sigma$ to the speed of light $c$. The product $\gamma\beta$ can then be given by $\gamma\beta=\sqrt{E_\Sigma^2/M_\Sigma^2-1}$. To evaluate the displaced vertex, we take into account the light neutrino masses with the minimal one being $10^{-4}$ eV.
In our analysis, $E_\Sigma$ is roughly taken to be $\sqrt{s}/2$ in the c.m. frame. Fig.~\ref{fig:lifetime} shows the decay length versus $M_N$ for $N_i~(i=1,2,3)$ and $\Omega=I$ at different collider energies. We find that in most cases the decay length is less than $\mathcal{O}(1)$ mm and the heavy leptons can be viewed as decaying promptly. Only for the cases of $N_1$ in NH and $N_3$ in IH, the typical decay length can be longer than 1 mm in relatively low mass region. In particular, for $M_N \sim 1$~TeV and $\sqrt{s}=30$ TeV, the decay length could be as large as 20~mm, which is sufficiently long to perform the displaced vertex searches~\cite{Jana:2020qzn,Sen:2021fha}.
If this clear difference is observed in future, it could serve as an indication to distinguish the heavy leptons and neutrino patterns.

%%%%%%%%%%%%%%%%%%%%%%%%%%%%%%%%
\section{Conclusion}
\label{sec:Con}
%%%%%%%%%%%%%%%%%%%%%%%%%%%%%%%%

We perform a study of search for heavy triplet leptons in Type III Seesaw mechanism at future high-energy and high-luminosity muon colliders.
The impact of up-to-date neutrino oscillation results is taken into account on neutrino mass model and the consequent lepton flavor signatures of heavy leptons in Type III Seesaw. According to the asymptotic behavior, the heavy lepton branching fractions exhibit ${\rm BR}(E^\pm\to \sum_\nu \overset{(-)}\nu W^\pm)\approx 50\%$ and ${\rm BR}(N\to \ell^\pm W^\mp)\approx {\rm BR}(E^\pm\to \ell^\pm Z)$ for $M_{E(N)}\gg M_W$. The three heavy leptons
in the Casas-Ibarra parametrization with diagonal unity matrix can be distinguished in terms of their decay modes.

We consider the pair production of charged triplet leptons $E^+ E^-$ through both $\mu^+\mu^-$ annihilation and VBF processes.
The charged triplet lepton as heavy as one-half of colliding energy can be probed. A $5\sigma$ significance can be reached through the $E^+E^-\to W^+W^-\nu\bar{\nu}$ process with the integrated luminosity below 0.1 ab$^{-1}$ for $\sqrt{s}=3$ TeV and 2 ab$^{-1}$ for $\sqrt{s}=10$ TeV.
The $E^+E^-\to ZZ\ell^+\ell^-$ channel can be further utilized to fully reconstruct the three triplet leptons and distinguish neutrino mass patterns. BR$(E^\pm\to Z\mu^\pm)$ below 10.7\% (6.4\%) can be reached for 5$\sigma$ (3$\sigma$) significance for $\sqrt{s}=3$ TeV and $\mathcal{L}=1~{\rm ab}^{-1}$.

The associated production $E^\pm N$ and the pair production of heavy neutrinos $NN$ are only induced by VBF processes and lead to rare LNV signature. We study the search potential of LNV processes at future high-energy muon collider with $\sqrt{s}=30$ TeV.
Through $E^\pm N$ ($NN$) production, ${\rm BR}(N\to \mu^{\pm}W^\mp)$ below 40.3\% (22.2\%) can be reached for 10 signal events and $M_{E(N)}<5$ TeV at muon collider with $\sqrt{s}=30$ TeV and $\mathcal{L}=90$ ${\rm ab}^{-1}$.

%###################################################################
%%%%%%%%%%%%%%%%%%%%%%%%
\acknowledgments
%%%%%%%%%%%%%%%%%%%%%%%%
We would like to thank Tao Han for very useful discussions.
T.L. is supported by the National Natural Science Foundation of China (Grant No. 11975129, 12035008) and ``the Fundamental Research Funds for the Central Universities'', Nankai University (Grants No. 63196013). C.Y.Y. is supported in part by the Grants No.~NSFC-11975130, No.~NSFC-12035008, No.~NSFC-12047533, the Helmholtz-OCPC International Postdoctoral Exchange Fellowship Program, the National Key Research and Development Program of China under Grant No.~2017YFA0402200 and the China Postdoctoral Science Foundation under Grant No.~2018M641621.
%%%%%%%%%%%%%%%%%%%%%%%%%%%%%%%%%%%%%%%%%%%%%%
%%%%%%%%%%%%%%%%%%%%%%%%%%%%%%%%%%%%%%%%%%%%%%
%%%%%%%%%%%%%%%%%%%%%%%%%%%%%%%%%%%%%
\bibliography{refs}

\end{document}